\documentclass[12pt,preprint]{aastex}

\shorttitle{3C279}
\shortauthors{Kartaltepe \& Balonek}

\begin{document}

\title{The Multiple Timescales of Optical Variability of the Blazar 3C 279 During the 2001-2002 Outburst}

\author{Jeyhan S. Kartaltepe\altaffilmark{1} \& Thomas J Balonek}

\affil{Department of Physics and Astronomy, Colgate University, 13 Oak Drive, Hamilton, NY 13346}
\email{jeyhan@ifa.hawaii.edu, tbalonek@mail.colgate.edu}

\altaffiltext{1}{Present Address: Institute for Astronomy, 2680 Woodlawn Dr., University of Hawaii, Honolulu, Hawaii, 96822}

\begin{abstract}
During 2001-2002 the optically violent variable (OVV) blazar 3C 279 underwent the most intense outburst seen during the entire fourteen year history that this quasar has been studied at Colgate University's Foggy Bottom Observatory (FBO).  This study concentrates on $\sim1600$ R-filter images taken during this period of activity. This data set includes twenty-nine nights of microvariability coverage. The outburst began in March 2001, after 3C 279 had faded to its faintest level, $R = 15.5$, in four years. The source reached its brightest level, $R = 12.5$, in the fourteen years of our study in August 2001, at which time it became unobservable due to its proximity to the Sun. Upon becoming observable again in mid-December 2001, 3C 279 fluctuated between $R=13.9$ and $R=14.7$, until a dramatic decrease in flux level in June-July 2002 brought the source back down to a level comparable to its pre-outburst state.  The source exhibited numerous week-long flares of approximately one magnitude during the outburst period.  Superposed on these flares were night-to-night variations of up to one half magnitude and intra-night microvariability of up to 0.13 magnitude in three hours. We use visual inspection of the light curve as well as numerical timescale analysis tools (the autocorrelation function, the structure function, and the power spectrum) to characterize the multiple timescales of variability ranging from 1.5 years to several hours.
\end{abstract}


\section{INTRODUCTION}

The blazar 3C 279 ($z = 0.538$), also known as 1253-055, is an optically violent variable (OVV) quasar representative of its class. It was the first quasar found to exhibit apparent superluminal motion (Whitney et al. 1971, Unwin et al. 1989), {\bfseries the first discovered to emit gamma rays}, by EGRET (Hartman et al. 1992), and is one of the brightest gamma ray sources in the sky (Kniffen et al. 1993, Hartman et al. 1996). 3C 279 is variable at all wavelengths of the spectrum and thus has become the focus of many multiwavelength campaigns (e.g., Netzer et al. 1994, Grandi et al. 1996, Wehrle et al. 1998). It is intensely variable at x-ray wavelengths (Lawson, McHardy, \& Marscher, 1999) and high-resolution VLBI radio maps show correlations between flare activity and components being ejected from the core (see, for example, Wehrle et al. 2001). Its highly polarized flux indicates that the radio through optical emission is likely to be dominated by synchrotron radiation (Hughes, Aller, \& Aller 1991, Maraschi et al. 1994) and it has been found to exhibit intense flux variability, particularly at shorter (optical through gamma ray) wavelengths (Webb et al. 1990, Edelson 1992, Stevens et al. 1994).

Optically, 3C 279 varies dramatically on many different timescales. Historically it has been seen to have a B magnitude range of $11.3 - 18.0$ (Netzer et al. 1994).  Studies of old photographic plates conducted after its discovery (Sandage \& Wyndham 1964) produced a light curve that dates back to 1929 (Eachus \& Liller 1975). This means that there are nearly seventy-five years of optical data for 3C 279 -- useful for long timescale analysis. This historical data also shows the first observed ``outburst" in 3C 279 in 1936-1937.  This is the classic outburst used for comparison with more recent outbursts.

The 2001-2002 outburst began in early 2001 when 3C 279 was the faintest it had been since 1997; by mid-2001 it had reached the brightest peak witnessed in the fourteen year optical monitoring program at Colgate University. The source returned to its pre-outburst flux by mid-2002. Our study of this outburst shows that it is one of the most intense, well-sampled outbursts of 3C 279. Even though it did not reach the peak ($B\sim11.3$) that the 1936-1937 outburst did, this outburst is much better sampled. Therefore, this data set is particularly valuable for studying the multiple timescale variations that occur within an outburst, including possible microvariability. Microvariability studies can help to place limits on current models of the origin of these small timescale variations and on the size of the emitting region (Mangalam \& Wiita 1993, Wagner \& Witzel 1995).  Microvariability has been seen to occur in many BL Lac objects and OVVs (see, for example, Miller, Carini \& Goodrich 1989 and Wagner \& Witell 1995), but 3C 279 itself has not been studied extensively for microvariability by previous studies even though some of these studies have found that 3C 279 can vary by more than one magnitude in one day (Webb et al. 1990, Sadun, Fajardo, \& Carini 1992). During one night of microvariability observations in 1989, Miller \& Noble (1996) found that 3C 279 underwent an event with a 0.15 magnitude amplitude in three hours.

In this paper, the term {\it outburst} will refer to long timescale variations on the order of a couple of years that last from a local minimum to the next local minimum. A period of sudden, increased intense activity within an outburst that lasts on the order of a few months is termed a {\it blaze}, since this activity appears to be characteristic of many blazars. {\it Flares} (\S3) are shorter events, with timescales on the order of a couple of weeks. There are also {\it ensembles of flares} that group together on the order of a couple of months. Note that flares and flare ensembles can be present within a blaze, superposed on the changing underlying brightness. {\it Intraday variability} will refer to smooth variations that occur over the course of a night and {\it microvariability} will refer to smaller fluctuations that occur on timescales of a few hours or less. This paper analyzes the different timescales of variability of 3C 279 during the 2001-2002 outburst, both qualitatively through visual inspection of the light curves, and quantitatively using several time analysis tools, using the well-sampled data obtained at the Foggy Bottom Observatory. In particular, we identify the typical timescales of variability and characterize the nature of its microvariability.

\section{OBSERVATIONS AND DATA REDUCTION}

Between 1989 and 2002, over 4600 images of 3C 279 on nearly 800 nights were taken at Colgate University's Foggy Bottom Observatory (FBO) using a Ferson sixteen-inch Newtonian/Cassegrain reflecting telescope equipped with a Photometrics PM3000 and Star I CCD system.  The images range in exposure time from one to five minutes (most images being two minutes) and include the B, V, R, and I filters, designed by Beckert (1989) to conform to the Johnson-Cousins system, though the majority of images were taken in the R filter. The images were reduced using standard IRAF packages and customized scripts written to facilitate the data handling for our system. 

The photometry for all of the images was calculated using the IRAF apphot package with a sixteen-arcsecond diameter aperture and a sky annulus with an inner diameter of twenty arcseconds and an outer diameter of forty arcseconds. As a comparison star, we used star 2 of Smith \& Balonek (1998), with a known magnitude of $R = 12.05 \pm 0.02$ (see Figure 1).  The color index for the comparison star is $V-R = 0.37$ (Smith \& Balonek 1998) and from our data the quasar is typically around $V-R = 0.50$. The color difference is therefore  $(V-R) \sim 0.13$. Given the near zero R color transformation coefficient of our filter set, the color correction for the quasar is insignificant for our data given the typical one-sigma error. All error bars are one-sigma and were calculated using the IRAF\footnote{IRAF is distributed by the National Optical Astronomy Observatories, which are operated by the Association of Universities for Research in Astronomy, Inc., under cooperative agreement with the National Science Foundation.} apphot package.  We thoroughly tested these error bars by measuring the scatter within a night for the quasar as well as within the entire dataset for several comparison stars.  We found that these error bars accurately represent this scatter, regardless of the magnitude of the source.

The fourteen year light curve is shown in Figure 2. Unless otherwise stated, all data points shown are nightly averages of typically half a dozen measurements taken within a half hour, except for nights when there is microvariability coverage, in which case the night is broken up into several groups of four to eight images that are averaged together. During the past fourteen years, 3C 279 has exhibited variability on many different timescales, ranging from intense outbursts that last about a year to microvariability on the scale of hours. Within this range there are also periods of activity on the scale of weeks to months.  There are gaps in the data every year between August and November when the quasar is not visible because of its proximity to the Sun. The faintest observed level of the source was in early 1990 at $R \sim 16.6$, and the brightest was in the summer of 2001 at $R \sim 12.4$.  In general, minima are in the range of $R \sim 15.5 - 16.5$ and maxima are in the range of $R ~ 13-14$. This data set will be fully analyzed in its historical context in another paper (T.J. Balonek et al., in preparation). 

The 2001-2002 outburst (see Figure 3) began around 2001.2 when 3C 279 was at its faintest level in five years, $R \sim 15.5$.  The outburst reached its peak in mid-August (2001.6) at which point it was at the brightest level ($R \sim 12.5$) observed in fourteen years of monitoring at FBO. 3C 279 was fading rapidly from this peak when it became unobservable due to its close proximity to the Sun. In December 2001, when the source became observable again, it had faded to $R \sim 14.4$. Throughout the spring of 2002, the source fluctuated between $R \sim 13.9$ and $R \sim 14.7$, unlike the large amplitude variations seen throughout 2001. In early summer the source began a quick decline back down to a flux level comparable with its pre-outburst state ($R \sim 15.6$). 

The light curve of the 2001-2002 outburst is particularly well sampled due to favorable weather conditions in the spring. This outburst is the most intensive outburst observed in this source at FBO, with an overall increase of over three magnitudes in less than five months. Superposed on this general rise are numerous flares that last about a week. Flares within this outburst, as well as within the general dataset, seem to follow the trend of having a steep rise followed by a slower decline. This variability profile is consistent with the optical emission of 3C 279 being dominated by synchrotron emission produced in the strong magnetic field of the relativistic jet. The largest observed increase in brightness is half a magnitude in one day.

\section{FLARES AND MICROVARIABILITY}

In this dataset, flares are seen to typically last between one and two weeks. In order to identify individual flares, Figures 4 and 11 show the light curves during 2001 and 2002, respectively. The various time periods discussed below are labeled individually in these figures. Several well-sampled flares are shown in this section with their respective nights of microvariability coverage. Shown in Table 1 are the individual measurements on the nights of microvariability coverage. 

\subsection{April 13 -- April 26, 2001}

Figure 5 shows the first well-sampled flare, or possibly two overlapping flares, lasting about two weeks, in which we obtained microvariability coverage during the outburst. The interpretation of this period of activity as two overlapping flares is supported by observations by Xie et al. (2002) from April 16 - 19. Between April 14 and 15, 3C 279 brightened by more than half a magnitude and then between April 19 and April 23 it brightened by an additional 0.5 magnitude to reach a peak of $R=13.6$ on April 23.  The quasar then decreased by 0.8 magnitudes over the next three days. The overall amplitude of this flare is nearly 1.5 magnitudes. From the microvariability coverage discussed below, there are fluctuations observed during this flare on timescales of hours.

\begin{description}

\item[\em {\it April 15}:\rm] This was the first night of microvariability coverage at FBO during the outburst. All of the scatter seen within the three and a half hours of observation is consistent with the error bars shown. The lack of variability within this night is interesting to note given that 3C 279 had brightened by half a magnitude in the previous 24 hours.

\item[\em {\it April 19}:\rm]	This night showed a possible decrease of $\sim 3\%$ in two hours followed by an increase of 12\% in brightness within about four hours of observations. It is also possible that there was a slight increase and then decrease in brightness during the first two hours, but with the intermittent sampling (due to observing other sources) this is inconclusive.

\item[\em{\it April 20}:\rm]	The light curve shows some smaller fluctuations on the order of 5\% within the total six hours of coverage. Given the particularly small error bars for this night, these fluctuations are believable, despite the sparse coverage. There was a steady rise beginning at 6 UT of about 5\% in less than one hour.

\end{description}

\subsection{April 26 -- May 3, 2001}

This flare (see Figure 6) began slowly but then rapidly sped up, increasing by 15\% between April 26th and 27th, and then by 45\% between the 27th and 28th. The source brightened by half a magnitude, or 58\% in one day right before the peak on April 29th. The overall rise for this flare was a full magnitude in just three days, or an average of 0.33 magnitude per day.  This corresponds to an overall increase in brightness of 150\%. The decline also began slowly and then the rate later increased. The overall decline was 0.48 magnitude, or a decrease of 55\% in four days. This is an average decline of 0.12 magnitude per day. Note that the rate of decline of this flare is significantly lower than the rate of increase. Since this flare is particularly well sampled (measurements on eight consecutive nights, including six nights of microvariability coverage) it is one of the best in our dataset and so is well suited for studying the behavior that is more difficult to see in the more complicated, less well sampled flares. 

\begin{description}
\item[\em{\it April 27}:\rm]	The coverage for this night possibly indicates a slight rise of about a tenth of a magnitude in six hours. This slope agrees well with the slope of the overall rise of the flare.

\item[\em{\it April 28}:\rm]	Despite an overall rise of 0.9 magnitude between April 27th and 29th, 3C 279 remained relatively constant during the five hours of observations on this night. This is similar to the behavior seen on April 15. Were it not for the microvariability coverage on this night, it would appear that the source was rising steadily. However, we know that this is not the case. This is an indication that the source does not increase steadily during a flare, but instead has variations within the rise itself. 

\item[\em{\it April 29}:\rm]	The peak of the flare was observed on this night and the light curve shows a decline of about four hundredths of a magnitude in three hours. The error bars are consistent with this general decline and so it is not likely that there are smaller fluctuations seen on this night. 

\item[\em{\it April 30}:\rm]	This night comes after the peak of the flare, but showed that the source rose slightly by about 6\% in two hours and then slowly decreased by 3\% in the next three hours. This night is also a good example of short timescale variability (on the order of a couple of hours) that differs from the longer timescale variability of the underlying flare.

\item[\em{\it May 1}:\rm]		This night showed a steady decline of 5\% in about two hours, followed by three hours of relatively constant brightness. All of the smaller timescale scatter within this night is consistent with the error bars.

\item[\em{\it May 2}:\rm]		The two hours of coverage for this night show no variation. Any scatter is consistent with the large error bars for these observations.

\end{description}

\subsection{May 7 -- May 21, 2001}

This two-week period of activity (see Figure 7) consists of many small timescale overlapping flares (lasting a few days). The overall brightness of 3C 279 during these two weeks did not change much, especially in comparison with the activity in April. The night-to-night variations during this period of activity are more erratic and the microvariability seen is greater than was the case for the other flares. There was an overall slight rise between May 7 and May 17 of $\sim0.2$ magnitude followed by a sudden decrease of about half a magnitude between the 17th and the 21st. These two weeks correspond to the central flat part of the outburst before the main peak (see Figure 4, 2001.4) when the source was not changing much.

\begin{description}
\item[\em{\it May 8}:\rm]		On this night, 3C 279 showed a steady decrease of 12\% in four and a half hours of monitoring and it is unlikely that there are any smaller fluctuations.

\item[\em{\it May 10}:\rm]	3C 279 was at a constant level of brightness ($R\sim13.7$) for the first hour of observations on this night and then slowly brightened by about 4\% over the next couple of hours. 

\item[\em{\it May 11}:\rm]	The two hours of observations for this night show that 3C 279 was constant. The scatter shown is consistent with the error bars.

\item[\em{\it May 14}:\rm]	The overall brightness of 3C 279 on this night was constant over the four hours of observations. It is possible that the smaller microvariability fluctuations seen here are real since the scatter is greater than one would expect from the particularly small error bars on this night. These variations are slightly less than 5\% over one-half to two hours.

\item[\em{\it May 15}:\rm] 	On this night 3C 279 faded by 0.13 magnitudes within three hours and then remained constant for about an hour. This is the largest small-timescale variation that we have seen within this microvariability data. It is not likely that there are any smaller fluctuations on shorter timescales within this night of data.

\item[\em{\it May 16}:\rm]	The limited coverage on this night of observations prevents the interpretation of small timescale variations, however a general rise of 7\% within about an hour is consistent with the data.

\item[\em{\it May 17}:\rm]	The night began with poor sky conditions, which improved as the night went on.  3C 279 dropped by 7\% in brightness over the course of three hours. Two nights later it was down to $R \sim 13.9$, a 32\% drop in two days.

\item[\em{\it May 20}:\rm]	On this night 3C 279 showed an overall decline of $\sim8\%$ within four hours. For the last two hours of observations the source appears to be relatively constant, though there might be some smaller timescale variations.

\item[\em{\it May 21}:\rm] 	The conditions for this night were poor as well, so the data show that 3C 279 was relatively constant for the three hours of observations given the error bars. 

\end{description}

\subsection{May 25 -- June 1, 2001}

The flare in Figure 8 is less intense and not as well sampled as the previous flares discussed. 3C 279 showed an overall change in brightness of about 0.4 magnitude during this one week. Since the source was not observed on every night during this week, it cannot be determined if there are smaller variations on the order of hours or days. 

\begin{description}

\item[\em{\it May 25 \& 28}:\rm] 	For the first two nights of microvariability coverage during this flare, 3C 279 was constant in brightness for two hours and one and a half hours on May 25 and May 28, respectively.

\item[\em{\it May 30}:\rm]	On this night 3C 279 dropped by about 2\% in brightness. The significance of this chance is inconclusive since it is not that much greater than the error bars. However, this decrease is consistent with the overall decline of the flare - and indication that the source may not have been constant during these observations.

\item[\em{\it May 31}:\rm]	3C 279 appeared to be constant in brightness for the three hours of observations on this night with the possibility of a brightening during the last hour of these observations. This is significant because this flare marks the period of least activity for 3C 279 since the outburst began and on this night it is at the faintest it has been in over a month. Immediately after this flare one of the most active periods for 3C 279 during this outburst begins.

\end{description}

\subsection{June 1 -- July 3, 2001}

This one month period (see Figure 9) of time reflects an intensely active phase coming after a relatively inactive phase for 3C 279. Within this month there is an overall increase of 0.8 magnitude in eight days and an overall decrease of a full magnitude over the course of the next three weeks. The steepest slope here is a decrease of 0.8 magnitude in four days at the end of the month. The good coverage during this flare ensemble allows us to see the overlapping nature of the flaring activity as well as witness several of the peaks themselves. 

The last nights of microvariability coverage were within this month. This coverage is for fewer hours than before since 3C 279 was lower in the sky during observations. After the minimum reached on July 2-3 ($R \sim 14.3$), 3C 279 began its swift increase toward its peak of $R=12.5$ in August (\S3.6), though this part of the outburst is not nearly as well sampled.

\begin{description}
\item[\em{\it June 7}:\rm]		This light curve shows a slight increase of about 5\% in two and a half hours. This night occurs during an overall increase in brightness for the flare that peaks the next day.

\item[\em{\it June 8}:\rm]		The quasar brightened by about 2\% during the first hour of observations and then faded suddenly by 6\%. The source then remained constant for the next half hour. This night occurs at the peak of the flare which faded by 15\% by the next night. Therefore it is likely that this sudden drop is real. 

\item[\em{\it June 9}:\rm]		On this night, 3C 279 remained at a constant brightness, with the possibility of a very slight decline, for two and a half hours of observations during an overall decline from the peak of the flare. 

\item[\em{\it June 13 \& 14}:\rm]	3C 279 remained at a constant brightness for an hour and a half on both of these nights. For June 14 it is possible that the data are consistent with a slight decline and then increase in brightness, though the scatter is consistent with the error bars.

\item[\em{\it June 18}:\rm]	The data for this night appear to be constant in brightness; however the scatter is not consistent with the error bars. It is likely that this light curve actually shows a slight decline followed by an increase in brightness and then possibly another decline. These small-scale fluctuations are on the timescale of about an hour.

\item[\em{\it June 20}:\rm]	On this night 3C 279 showed an increase in brightness of about 8\% in less than two hours.

\end{description}

Although less well sampled, there is another flare in this ensemble lasting from June 21 -- July 3. This flare shows a slow rise that peaks on the 26th and then a very rapid decline of $\sim0.9$ magnitudes over the next six days. This rapid decline is atypical of the type of behavior that we have witnessed in other flares.

\subsection{July 3 -- August 16, 2001}

The light curve during this time period (Figure 10) is less well sampled since it comes at the end of 3C 279's observing season. There was at least one flare during this period, a well-sampled one between July 3 and July 12 (2001.50 -- 2001.53). Overall, there was an increase of two magnitudes during this two-month period. One thing to note about these variations is that they are smaller fluctuations superposed on top of the gradual rise toward the peak of the outburst (the brightest peak seen in the 14-year dataset) that is observed on August 5, 2001, i.e., 3C 279 did not vary smoothly toward the peak of the outburst. Due to the sparseness of observations during this month, it is unlikely that the real peak of the outburst was on this date. Instead, it was probably reached sometime in the next couple of days.  The measurements after this night show 3C 279 beginning its decline from this peak and no smaller timescale variations are discernable.

\subsection{2002}

By the time 3C 279 became observable again in mid-December of 2001, it had faded to $R=14.3$, a full magnitude drop since the last measurement nearly three months prior. The behavior during 2002 (see Figures 3 and 11) was relatively stable compared to that during 2001. There are about seven easily distinguishable flares labeled $a-g$ in Figure 11.  However, most of these are not well sampled so it is likely that they actually represent several overlapping flares rather than individual flares. The first of these, $a$ (Dec 12 -- Jan 4), has a 0.3 magnitude amplitude and lasted for nearly one month. Due to the sampling, it is not clear whether or not there were any smaller variations. Flare $b$ (Jan 18 -- Feb 10) also lasted for almost a month and had an amplitude of 0.35 magnitude. This flare was slightly better sampled though and showed a rather steep increase (0.35 magnitude in 5 days) followed by a shallow decline (0.3 magnitude in 14 days).

Flare $c$ (Feb 10 -- 18) was much smaller in both duration and amplitude. It lasted for one week and had a 0.2 magnitude amplitude. The next flare, $d$ (Feb 18 -- Mar 14), lasted for nearly one month and had a 0.4 magnitude increase in two weeks followed by a 0.7 magnitude drop over the next ten days. Unlike many of the flares seen before, this flare had a steeper decline than rise, though the sampling makes it difficult to see variations on smaller timescales.  Flare $e$ (Mar 17 -- 28), though undersampled, lasted for about two weeks and has a half magnitude amplitude. Flare f (Apr 5 -- May 1) actually appears to be a group of at least three flares, but these are not possible to distinguish individually. The entire event lasts for about one month and shows a half magnitude rise over three weeks followed by a 0.7 magnitude drop over the next twelve days.  

The next event, $g$ (May 4 -- Jun 18), appears to actually be two overlapping flares. The entire event lasted for about a month and a half and has a 0.7 magnitude amplitude. 3C 279 reaches a peak of $R\sim14$ before beginning a rapid decline down to $R\sim15.5$ over the next two months. After this flare, no other complete flares are discernable. This level is back down to the pre-outburst level at the beginning of 2001 and so the 2001-2002 outburst is clearly over. Preliminary reductions indicate that 3C 279 has continued to drop since then and as of late-May 2003 had faded to a minimum of $R\sim17$, the faintest we have seen in our fourteen year study.

\section{TIME SERIES ANALYSIS}

Visual inspection of the light curves in \S3 yielded several characteristic timescales of variability. These timescales are on the order of hours (microvariability), days-weeks (flares), months (flare ensembles and blazes) and years (outbursts). In this section we employ several numerical techniques as another way to look for these variability timescales. The significance of the timescales found with these numerical techniques will be discussed in the next section and compared to those found by visual inspection.

\subsection{Autocorrelation Function}

One quantitative method for understanding the various timescales of variability is to look at the autocorrelation function (ACF) of the flux density of 3C 279. The cross correlation function for discrete, unevenly spaced data (DCF: Edelson \& Krolik, 1988) is given by: 

\begin{displaymath}
DCF(\tau) = \frac{1}{M}\frac{(g_i-\overline{g})(h_j-\overline{h})}{\sqrt{(\sigma^2_g-e^2_g)(\sigma^2_h-e^2_h)}},
\end{displaymath}

\noindent
where $g$ and $h$ are the two functions (which for an ACF are the same function) and $\overline{g}$ and $\overline{h}$ are the average values for the two functions, $M$ is the number of pairs (i,j) of points between the two functions, $\sigma$  is the standard deviation, and $e$ is the measurement error. 

A Fortran code provided by Jorstad (2003, private communication) was used to calculate the ACF in four-day bins for the entire fourteen-year dataset (see Figure 12). The characteristic timescale of variability is given by the width of the ACF near zero time lag (Netzer et al. 1996). For 3C 279, the width of the ACF is eight days. This function shows a large broad peak at $\sim550$ days, a smaller, narrow peak at $\sim360$ days, and a strong peak at $\sim100$ days. An ACF in one-day bins shows the higher resolution central portion of the graph (see Figure 13) to have additional structure, including strong peaks at approximately 40 and 7 days. The ACF for just 2001-2002 (Figure 14) also shows this characteristic timescale of 8 days as well as a peak at 45 days.

The ACF was also calculated using the individual measurements with different bins during the 2001-2002 outburst to look for small timescale variations. Figure 15 shows the ACF for one-hour bins. The width is equal to eight hours and there is an additional peak at 5 hours. Since our largest data samples are typically between 5 and 8 hours in duration, it is possible that these two timescales are simply aliases of the data sampling. There are also pairs of peaks at one day increments, most likely due to the data sampling interval of $\sim 1$ day. 

For the ACF with ten minutes bins (Figure 16) there are peaks at 1 day and at 20 hours, both of which are due to the sampling interval. The width of the function is one hour and there is a large peak at $\sim5.2$ hours. Figure 17 shows the ACF for all of the microvariability data using two-minute bins. The width of this peak is 2.3 hours and there are additional peaks at 1.5, 3, 5, and 6 hours. The three and six hour timescales are both multiples of the 1.5-hour timescale, so 1.5, 2.3 and 5.0 hours are the three significant timescales from this ACF. It is interesting to note that the aliasing timescales of $\sim5$ hours and 1 day were found in multiple bins of outburst data. 

\subsection{Power Spectrum}

The traditional method of timescale analysis is to compute power spectra for various parts of the dataset.  The equation for the power spectrum is given by (Scargle 1981):

\begin{displaymath}
PSD = \frac{1}{N}\left[\sum^N_{j=1} f(t_j)\:e^{-i\:w\:t_j} \right]^2,
\end{displaymath}

\noindent
where $N$ is the number of data points and $f(t)$ is the function (in this case the flux density of the quasar).  We adopted the CLEAN method to calculate the power spectra since it is particularly useful for handling unequally spaced data (Roberts, Lehar \& Dreher 1987) and we use a code provided by Jorstad (2003, private communication). Figure 18 shows a sample power spectrum for the entire fourteen-year dataset. Note that there are clumps of peaks at 540, 345, 200, 100 and 50 days. Using an approach similar to that for the autocorrelation function, we calculated power spectra for different subsets of the data and found timescales on the order of 1.5 years, 1 year, 100 days, 50 days, 10-20 days and 1 day. 

\subsection{Structure Function}
A third approach to timescale analysis is the structure function, discussed fully by Simonetti, Cordes, \& Heeschen (1985).  The structure function (SF) is particularly useful for analyzing non-periodic data with multiple variation timescales and is given by:

\begin{displaymath}
SF(t)=\left\langle \left[f(t)-f(t+dt)\right]^2\right\rangle,
\end{displaymath}

\noindent
where $f(t)$ is the given function, in this case, the flux density of the quasar. A break in the curve of a log-log plot of the structure function provides information about the variation timescales. 

Using a code provided by Jorstad (2003, private communication), the SF for the data from 2001-2002 was calculated and is shown in Figure 19.  The plot shows a turnover at 0.4 years and possible breaks at 46 and 14 days. The turnover at 0.4 years is caused by the rise time of the large outburst being roughly equal to the 0.4 years of data in 2001 before 3C 279 becomes unobservable. Right before this it had reached its peak brightness and began a decline. 

After becoming observable again, 3C 279 had decreased dramatically, as indicated by the turnover in the SF. Because of the coincidence of 3C 279 reaching its peak just before becoming unobservable, it is uncertain whether 0.4 years represents a valid variability timescale. It is not possible to know whether the peak witnessed in 2001 was the highest peak that 3C 279 reached, though the rapid decline that began right after this peak is evidence for that. The fact that 3C 279 was significantly fainter when observed three months later indicates that the break must be before 0.6 years. However, by looking at the SF for all the data, we can get a more accurate timescale that is not solely based on one outburst. Figure 20 shows the SF of the entire data set. The general shape of the SF appears to be the same, but now the turnover occurs at 0.3 years with possible breaks at 40 and 14 days. The fact that the turnover occurs sooner is due to the fact that other outbursts in the data set are shorter in duration than the 2001-2002 event.

Since the SF in each of these cases is dominated by the presence of extremely intensive outbursts, it is difficult to tell whether or not the other breaks accurately represent intrinsic timescales of variability. By looking at the SF for a subset of data when the quasar was relatively stable, it is possible to look for these smaller scale variations without having them drowned out by outbursts. Figure 21 shows the SF for the relatively stable 2002.  Since there are fewer data points in 2002, the scatter is larger. However, there is an apparent break at 7 days and possibly one at 41 days as well. These two timescales agree with the breaks found for the two previous SFs, with the 7 day one being much more apparent and the 41 day one being less clear.

\section{DISCUSSION AND CONCLUSION}
\subsection{Numerical Techniques}

Numerous variability timescales were found using the three numerical techniques discussed in \S4. The next step is to compare these timescales to our visual inspection of the light curve of 3C 279 to see if any of them represent valid variability timescales. Table 2 summarizes the results of these techniques. Both the ACF and PS found a timescale of one year. The only thing that repeats every year in the data is the annual gap due to 3C 279 becoming unobservable; therefore it is likely that this timescale is simply an artifact of the data-sampling interval. The ACF and PS also found a distinct one day timescale in the data that is also most likely due to data sampling, since measurements are usually taken about one day apart.

Both the ACF and the PS found a timescale of $\sim550$ days or 1.5 years. Outbursts correspond well to this variability timescale. The 2001-2002 outburst for example, lasts for 1.4 years. Other outbursts in the light curve last longer or shorter than this. One timescale found by all three numerical methods is the 100-day timescale. At first this was rather surprising, since there is not any obvious activity that lasts for 100 days. However, the duration of the intense period of activity, which we term a blaze, typically lasts around 100 days. An example of this is the activity seen during all of 2001. This activity was very intense and quickly rose to a very high peak. This type of activity is also clearly seen in early 1989, 1991, 1992 and 1998. 

All three methods also found the typical timescale of a flare (such as the ones discussed in \S3.1 - \S3.4) to be between seven and ten days. This is the most clearly defined timescale found and is the characteristic timescale found from the ACF. 3C 279 exhibits this type of activity at all times, whether bright or faint, during a blaze, or during relatively quiescent periods of time. This is an indication that the flare timescale is a characteristic of the source itself and therefore must be physically significant. All three methods also found the timescale of a flare ensemble (such as that seen in \S3.5 and in all of 2002 (\S3.7)) to be between 40 and 50 days.  

The ACF also found smaller timescale variations (8, 5, 2.3 and 1.5 hours) but since neither of the other two methods found them it is difficult to determine whether or not these timescales are accurate. It is likely that both the 8 and 5 hour timescales are due to data sampling. The 2.3 and 1.5 hour timescales likely represent the hour(s) long variations seen in the microvariability data. These timescales give some confidence to the microvariability results found from visual inspection in \S3. It is also possible, however, that these small timescales are caused by the length of many of the microvariablity observations (those on the order of a couple hours). Further tests are needed to confirm these microvariability timescales.

\subsection{Nature of the Variability}

Our data contains twenty-nine nights of microvariability coverage during the 2001-2002 outburst. Table 3 shows a summary of the results found for each of these nights. The date for each night, average error bar, overall slope, and multiple slopes for the cases where it changes sign over the course of the night are shown for comparison. On many of these nights significant microvariability is seen on the timescales of a few hours. The largest slopes seen are on May 15, with a decrease of 0.13 magnitude in three hours and on May 16 with an increase of 0.07 magnitude within an hour. The smallest slopes are negligible, consistent with 3C 279 being constant in brightness. 

The types of variations seen on these timescales fit into two different categories. There is intraday variability, smooth changes over the course of a night (such as May 15), which follow the general trend of a flare. There are also smaller fluctuations within a night that represent structure superposed on an underlying flare, such as the 0.05 magnitude fluctuations seen on May 14, or the lack of variability during the rise of a flare, such as on April 15 and April 28. It is worth noting, however, that in all of the nights of microvariability coverage, we never see the rapid large amplitude microvariability that is observed in other sources, such as BL Lacertae, which exhibits well defined events of more than half a magnitude within five hours (see, for example, Bloom et al. 1997). It is also notable that our observations did not show the type of microvariability, a complete event of 0.15 magnitude amplitude in three hours with substructure, seen in 3C 279 by Miller \& Noble (1996). Since all of our microvariability coverage took place when 3C 279 was relatively bright, it is unclear how the observed microvariability relates to the overall luminosity of the source.

From the analysis of the 2001-2002 outburst, we have found several overlapping timescales of variability.  Because of the overlapping nature of these different timescales, it is important to note the differences in behavior seen in 2001 and 2002. While throughout 2002 3C 279 exhibited several flare ensembles and variability on the order of weeks to months, the base level remained about the same. These fluctuations did not diverge greatly from the average of R~14.2.  During 2001, 3C 279 displayed variations on these same timescales (flares and flare ensembles), but they are superposed on an overall steep increase of three magnitudes in about five months. This behavior is radically different from that seen during 2002 and is what we have characterized as a blaze.

The 2001-2002 outburst is the most intensive and best-sampled outburst observed in 3C 279 at Colgate University. This dataset is therefore ideal to use to study the nature of the various variability timescales that are characteristic of 3C 279's history of optically violent variable behavior. We have characterized the various timescales through visual inspection and confirmed them through three different numerical techniques. These timescales range from 1.5 years to several hours. Significant microvariability has been observed in 3C 279 on the timescale of a few hours and it has therefore been shown that individual flares do not increase or decrease in brightness steadily, but there is structure within these changes on the order of the microvariability timescale. 

\acknowledgments

We would like to thank S. Jorstad for providing the programs to calculate the ACF, PS, and SF numerical techniques. Also, we would like to thank NASA/NY Space Grant, the Colgate University Alumni Memorial Fellowship \& Division of Natural Sciences and Mathematics for funding some of the research for this project, as well as the multitude of students who have observed 3C 279 at Colgate University over the past fourteen years.

%figure1
\begin{figure}
%\epsscale{0.25}
\plotone{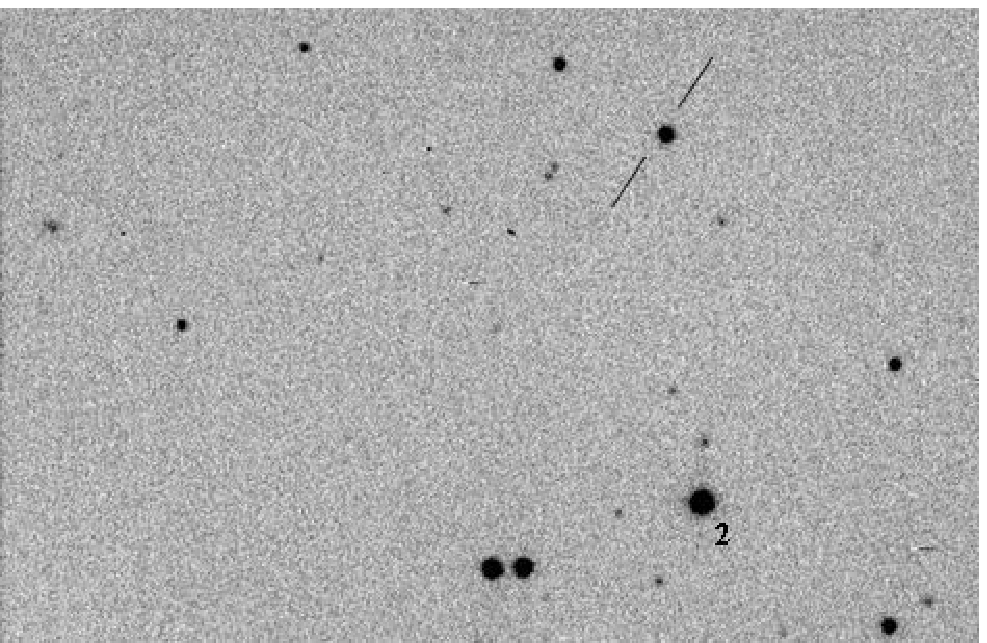}
\caption{Two-minute R-filter image of the 3C 279 field taken at FBO. 3C 279 is marked and the main comparison star is labeled number 2. In this image, north is up and east is to the left. The field of view is 5 by 8 arcminutes.}
\end{figure}

%figure2
\begin{figure}
\includegraphics[angle=90,width=5.9in]{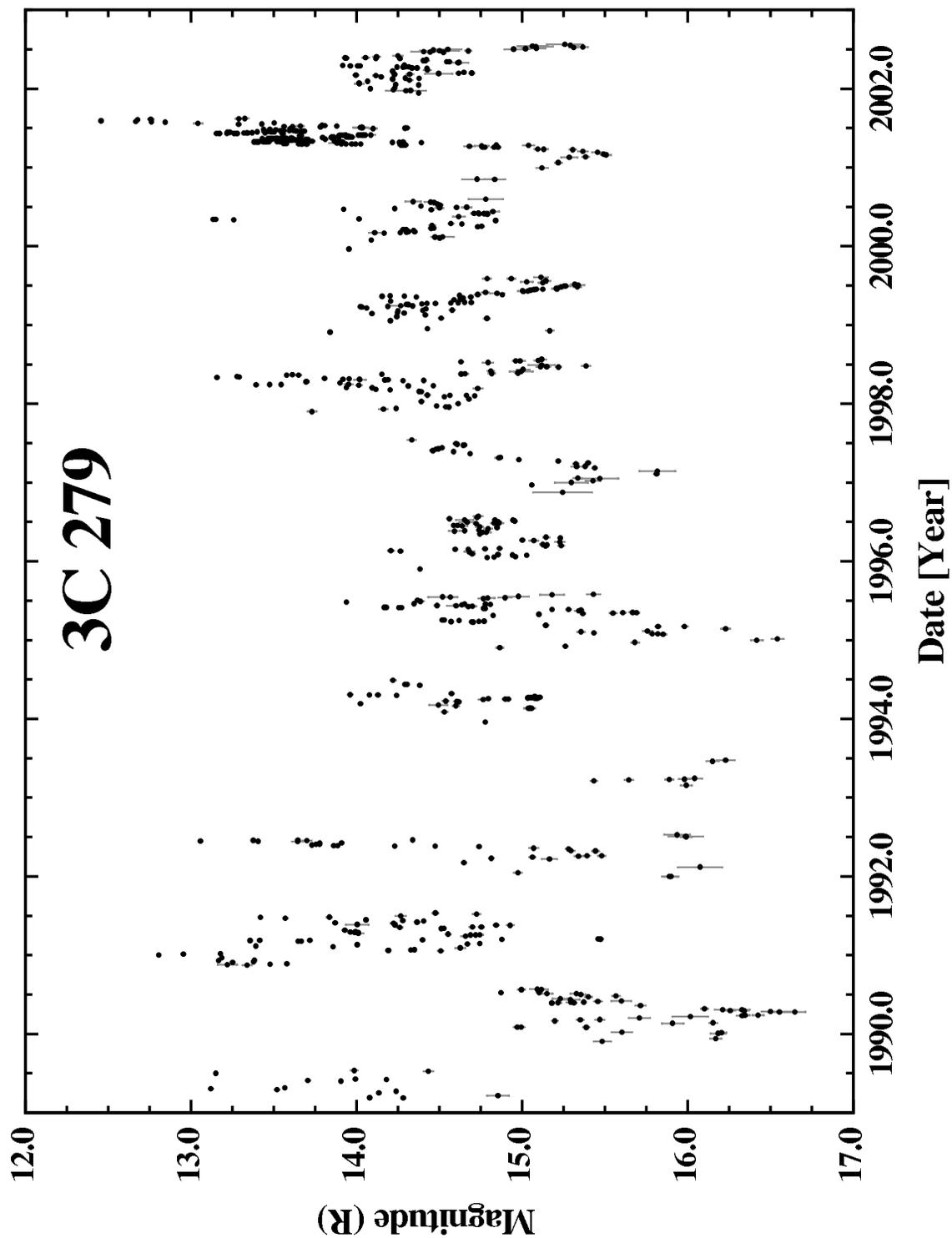}
\caption{Fourteen-year optical light curve of 3C 279. All data were taken in the R filter at FBO. All points are nightly averages, except for nights with microvariability coverage that are broken into groups of 4-8 images. Note the minimum in 1990 and the maximum in 2001. }
\end{figure}

%figure3
\begin{figure}
\includegraphics[angle=90,width=5.8in]{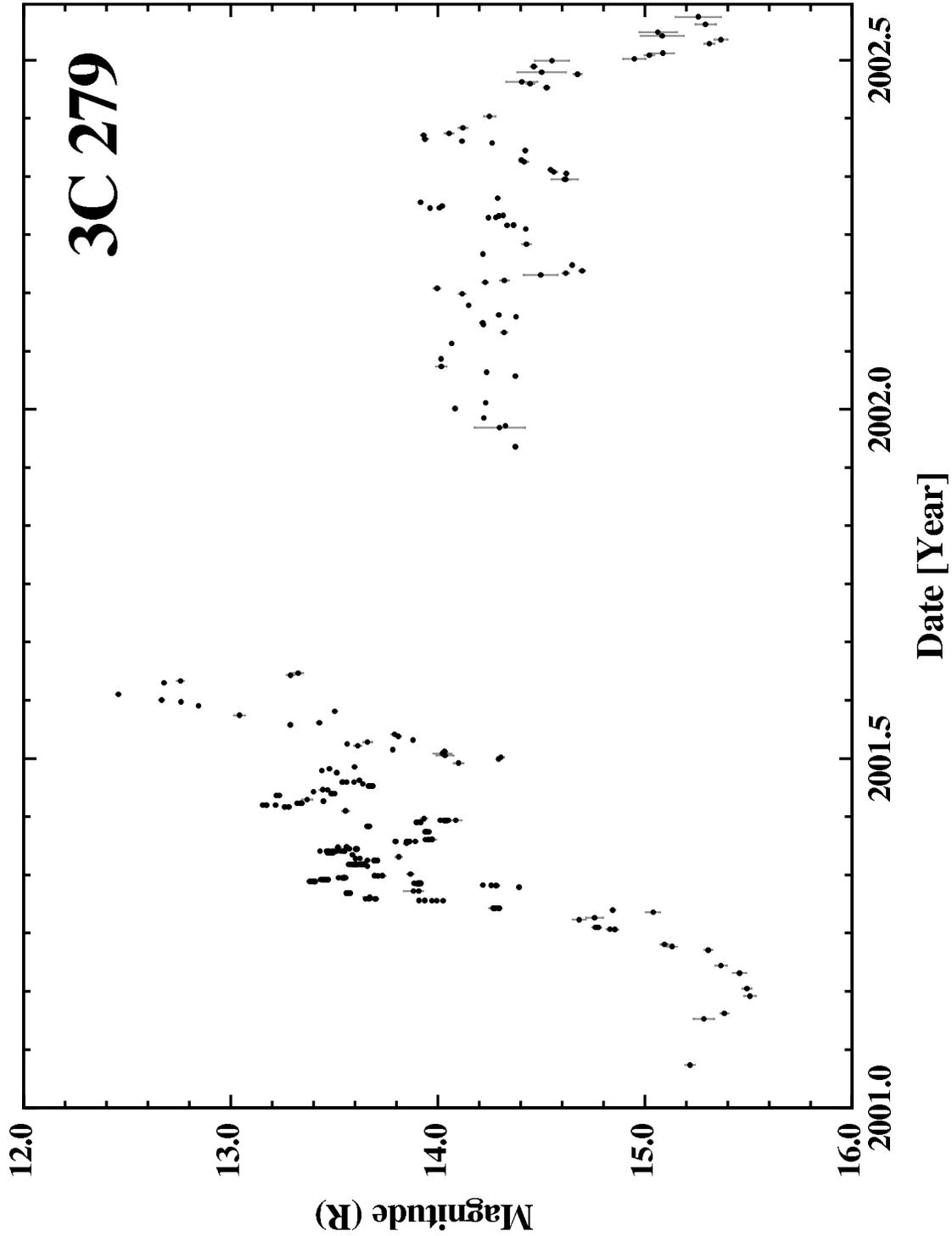}
\caption{Light curve of the 2001-2002 optical outburst of 3C 279. All points are nightly averages, except for nights with microvariability coverage that are broken into groups of $4-8$ images. Each horizontal axis tick corresponds to approximately one month of time.}
\end{figure}

%figure4
\begin{figure}
\includegraphics[angle=90,width=5.9in]{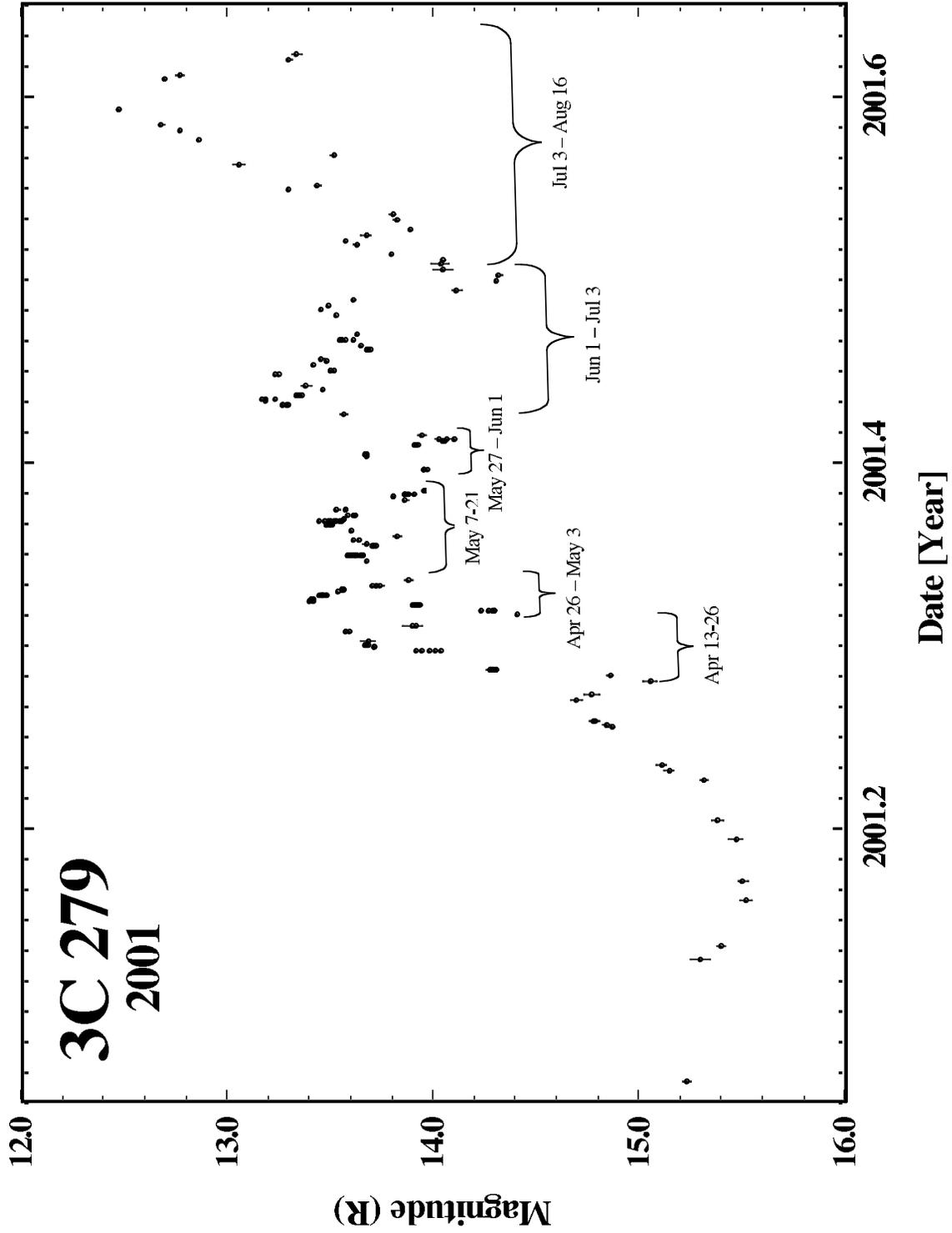}
\caption{Light curve of 3C 279 for 2001 only. The flares discussed in \S3 are individually labeled.}
\end{figure}

%figure5
\begin{figure}
\epsscale{0.82}
\plotone{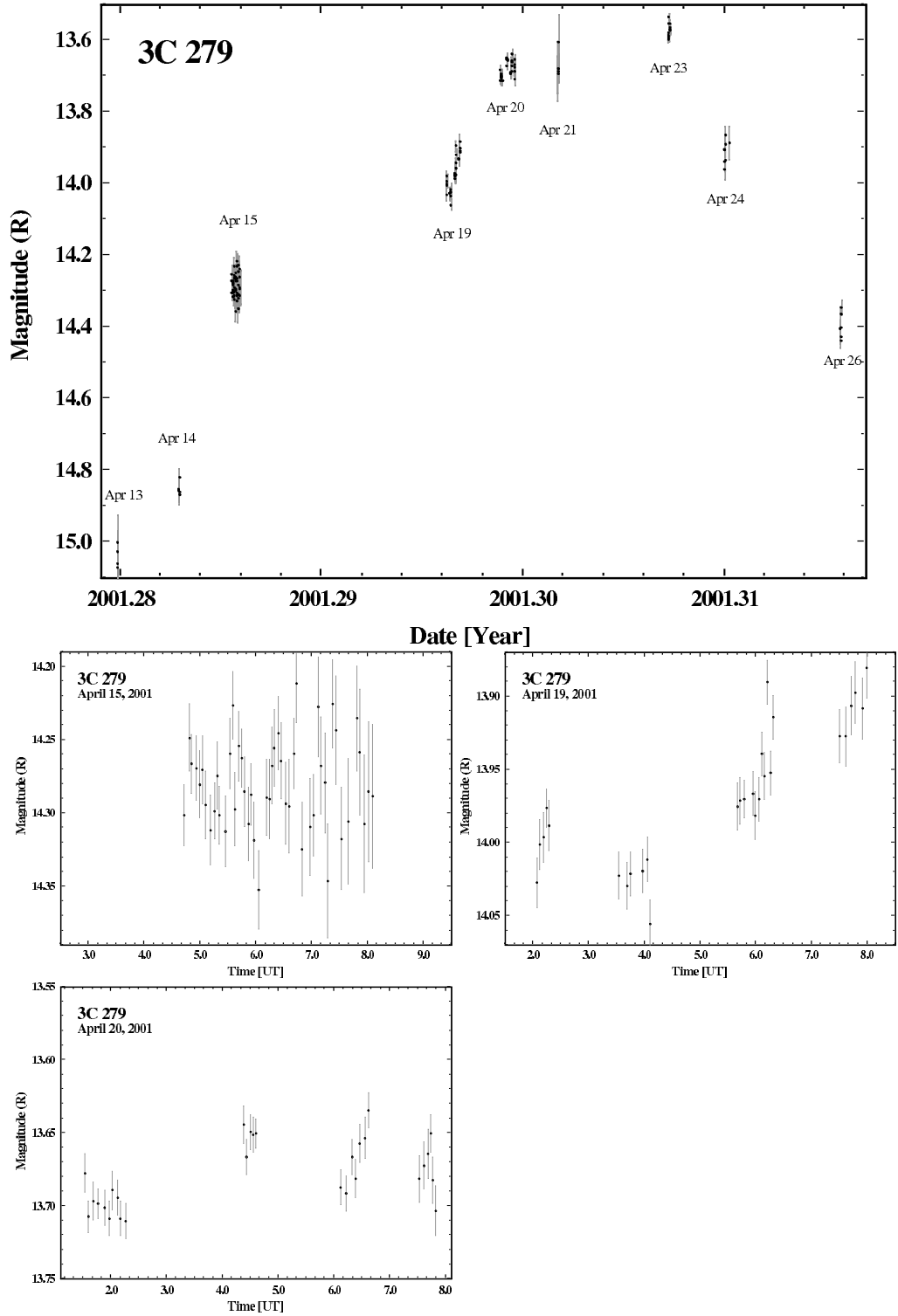}
\caption{Light curve of an individual flare (April 13-26) for 3C 279 during the 2001-2002 outburst and the corresponding nights of microvariability coverage. Each data point shown is an individual measurement and all error bars are 1$\sigma$.}
\end{figure}

%figure6
\begin{figure}
\epsscale{0.82}
\plotone{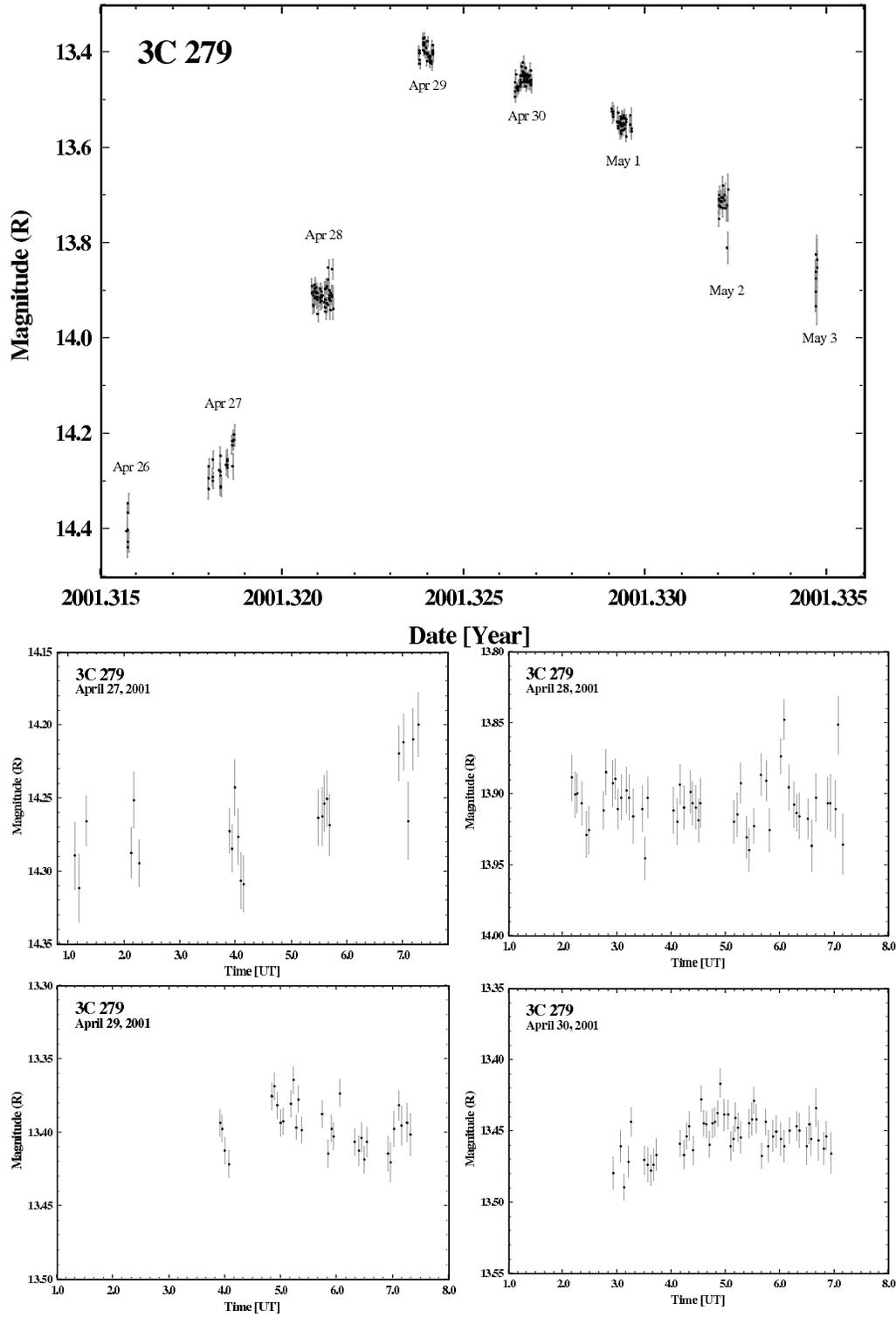}
\caption{Light curve of a flare ensemble (April 26-May 3) for 3C 279 during the 2001-2002 outburst and the corresponding nights of microvariability coverage. Each data point shown is an individual measurement and all error bars are 1$\sigma$.}
\end{figure}

\addtocounter{figure}{-1}

%figure6b
\begin{figure}
%\figurenum{6 -- cont.}
\epsscale{0.82}
\plotone{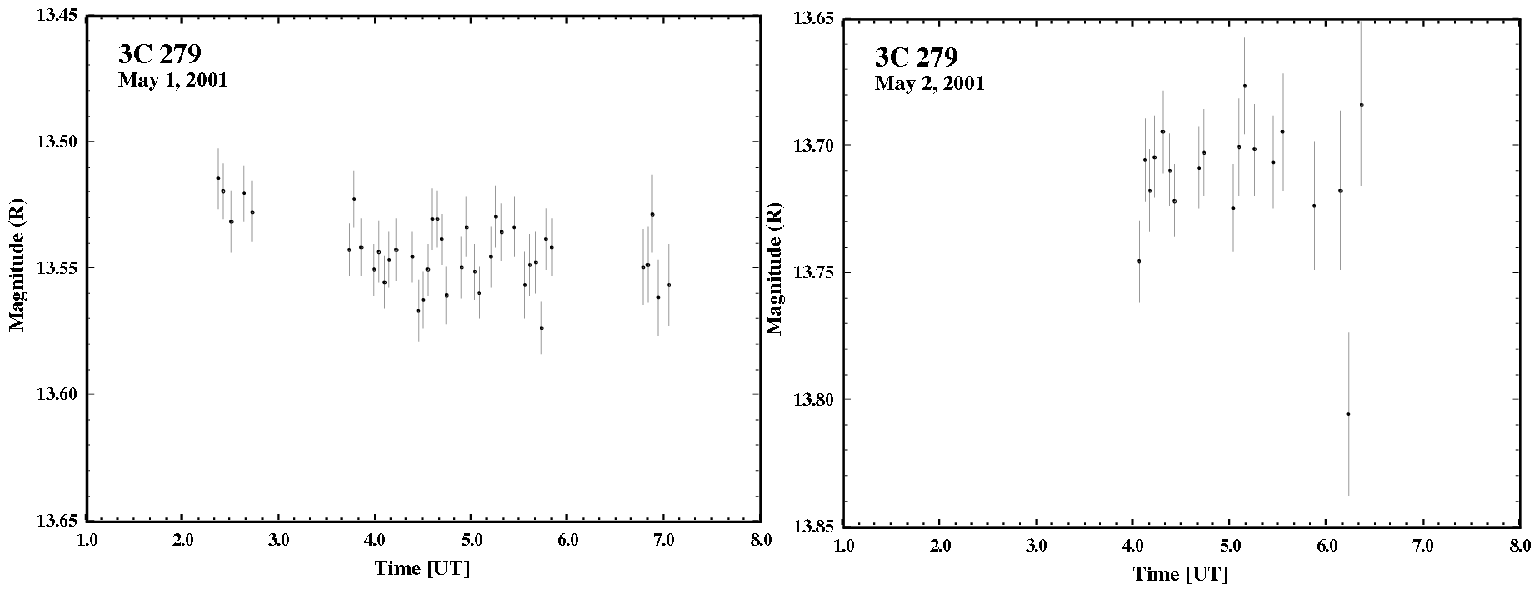}
\caption{continued}
\end{figure}

%figure7
\begin{figure}
\epsscale{0.80}
\plotone{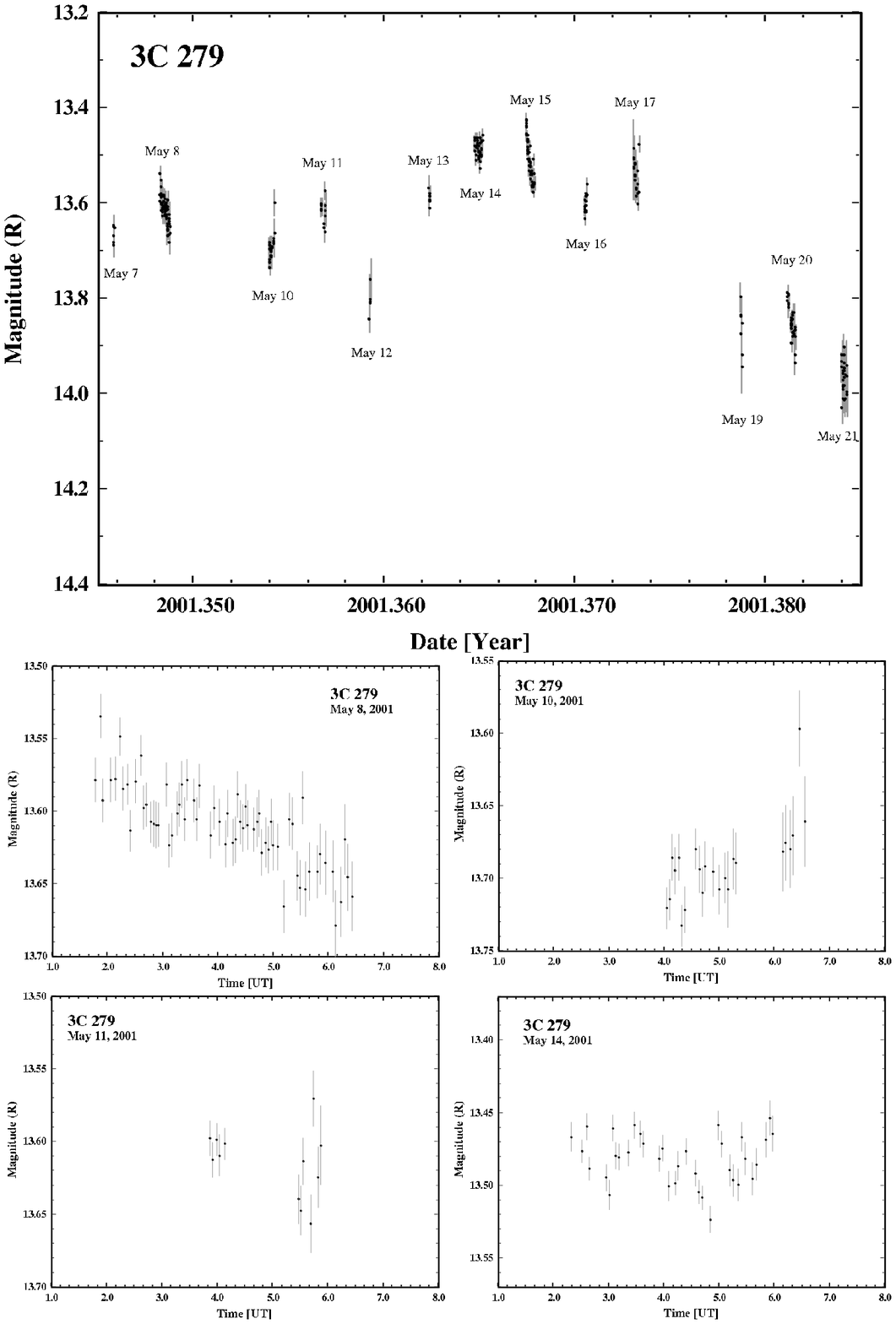}
\caption{Light curve of a flare ensemble for 3C 279 (May 7-21) during the 2001-2002 outburst and the corresponding nights of microvariability coverage. Each data point shown is an individual measurement and all error bars are 1$\sigma$.}
\end{figure}

\addtocounter{figure}{-1}

%figure7b
\begin{figure}
%\figurenum{7 -- cont.}
\epsscale{0.80}
\plotone{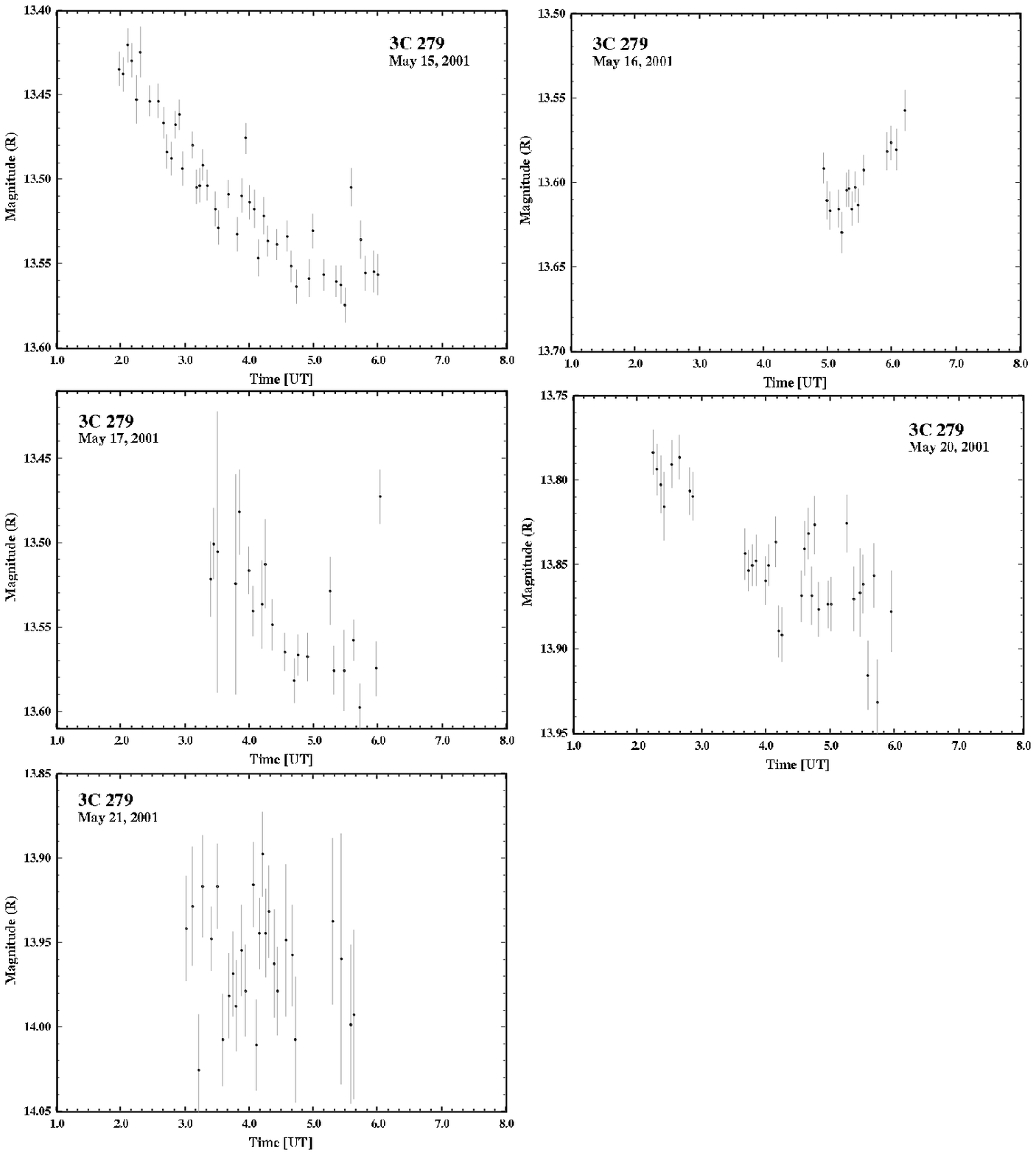}
\caption{continued}
\end{figure}

%figure8
\begin{figure}
\epsscale{0.80}
\plotone{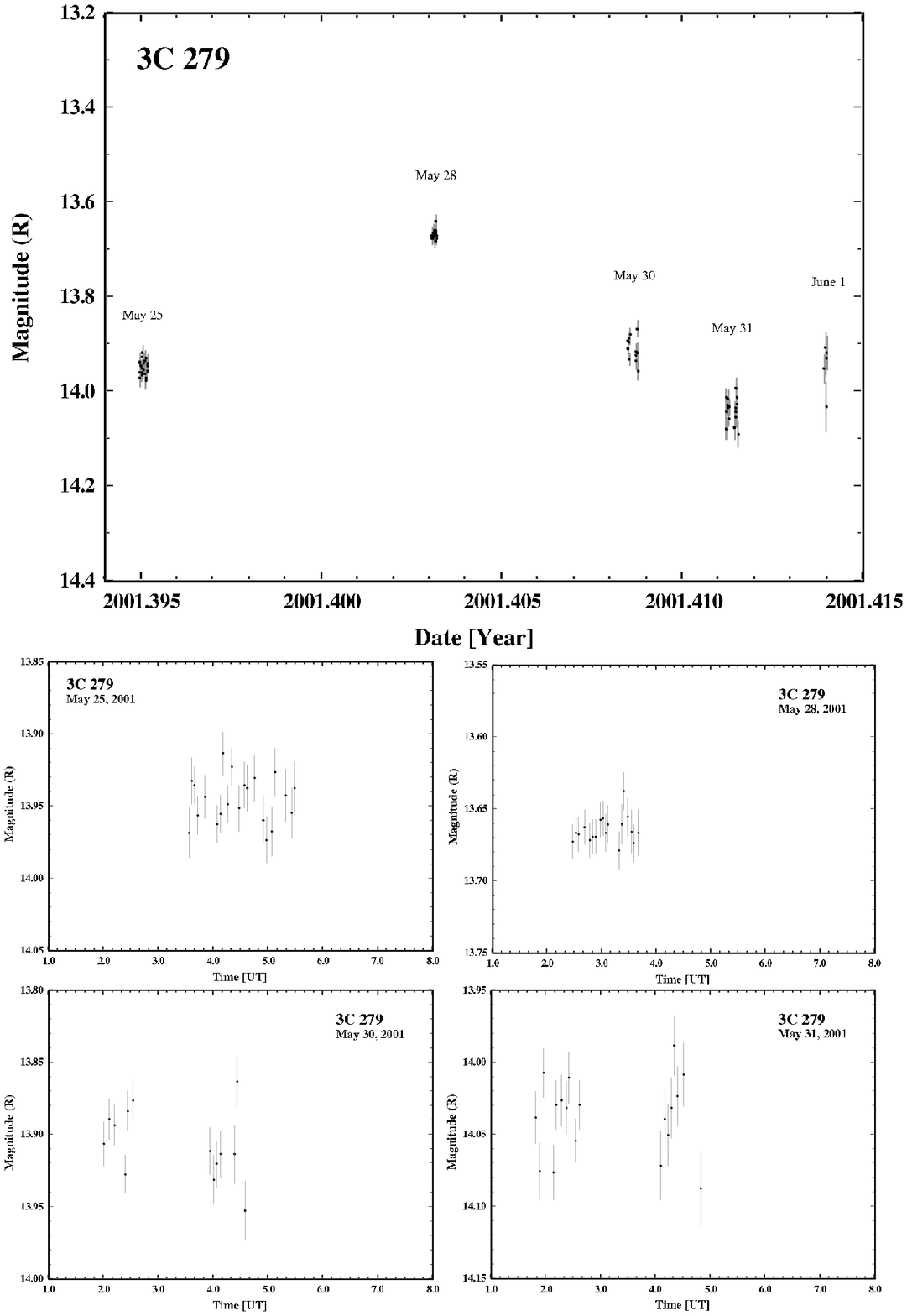}
\caption{Light curve of an individual flare for 3C 279 (May 25-June1) during the 2001-2002 outburst and the corresponding nights of microvariability coverage. Each data point shown is an individual measurement and all error bars are  1$\sigma$.}
\end{figure}

%figure9
\begin{figure}
\epsscale{0.79}
\plotone{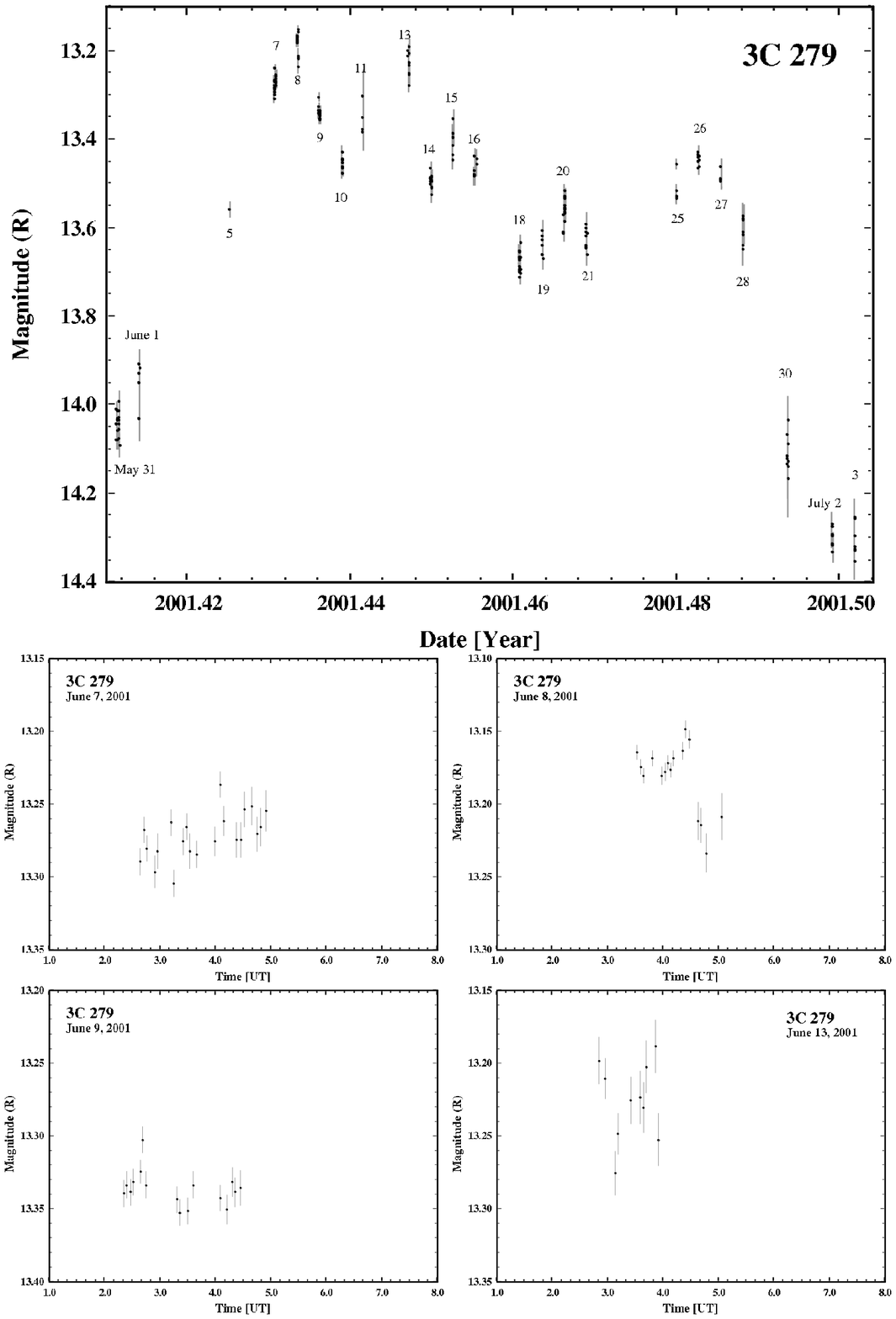}
\caption{Light curve of a flare ensemble for 3C 279 (May 31 - July 3) during the 2001-2002 outburst and the corresponding nights of microvariability coverage. Each data point shown is an individual measurement and all error bars are  1$\sigma$.}
\end{figure}

\addtocounter{figure}{-1}

%figure9b
\begin{figure}
%\figurenum{9 -- cont.}
\epsscale{0.79}
\plotone{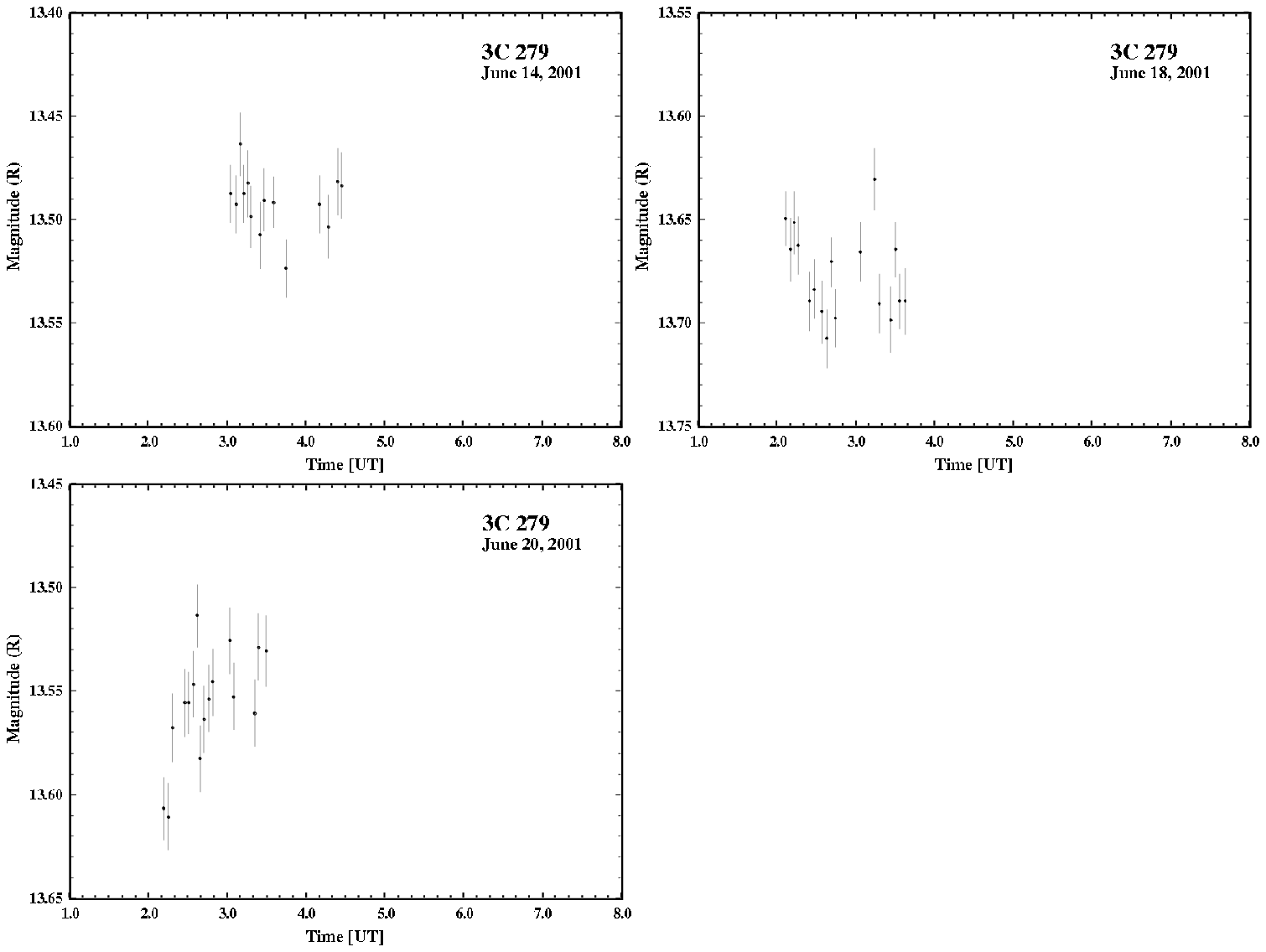}
\caption{continued}
\end{figure}

%figure10
\begin{figure}
\includegraphics[angle=90,width=5.8in]{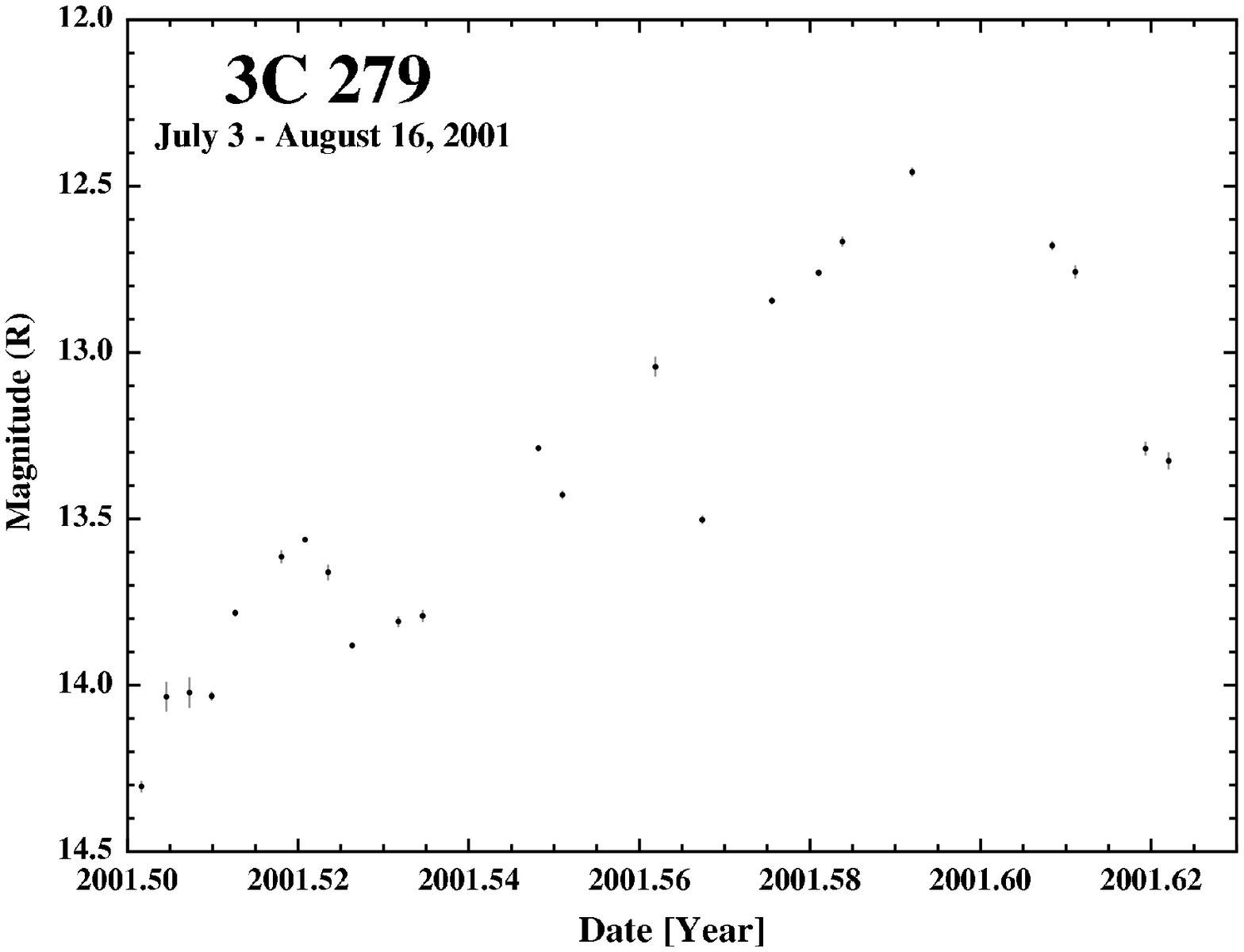}
\caption{Light curve of 3C 279 during the summer of 2001, from July 3 - August 16. These data points are nightly averages. The peak observed on August 5 is the brightest seen in 14 years of monitoring.}
\end{figure}

%figure11
\begin{figure}
\includegraphics[angle=90,width=5.8in]{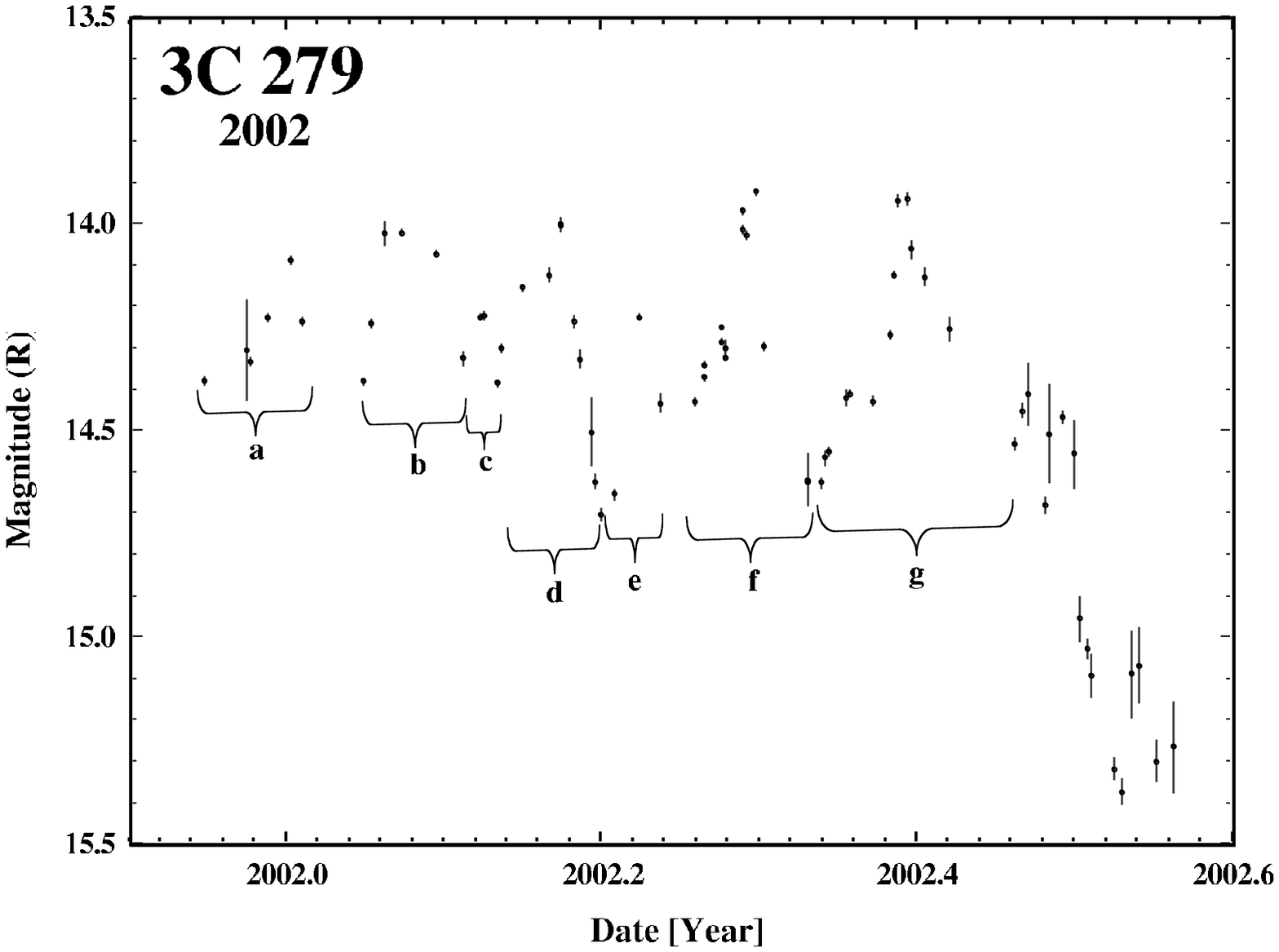}
\caption{: Light curve of 3C 279 during 2002. The flares discussed in \S3.7 are labeled $a-g$. The dates for each of the flares are: $a$ (Dec 12 - Jan 4), $b$ (Jan 18 - Feb 10), $c$ (Feb 10 - 18), d (Feb 18 - Mar 14), $e$ (Mar 17 - 28), $f$ (Apr 5 - May 1), and $g$ (May 4 - Jun 18).}
\end{figure}

%figure12
\begin{figure}
\epsscale{1.00}
\plotone{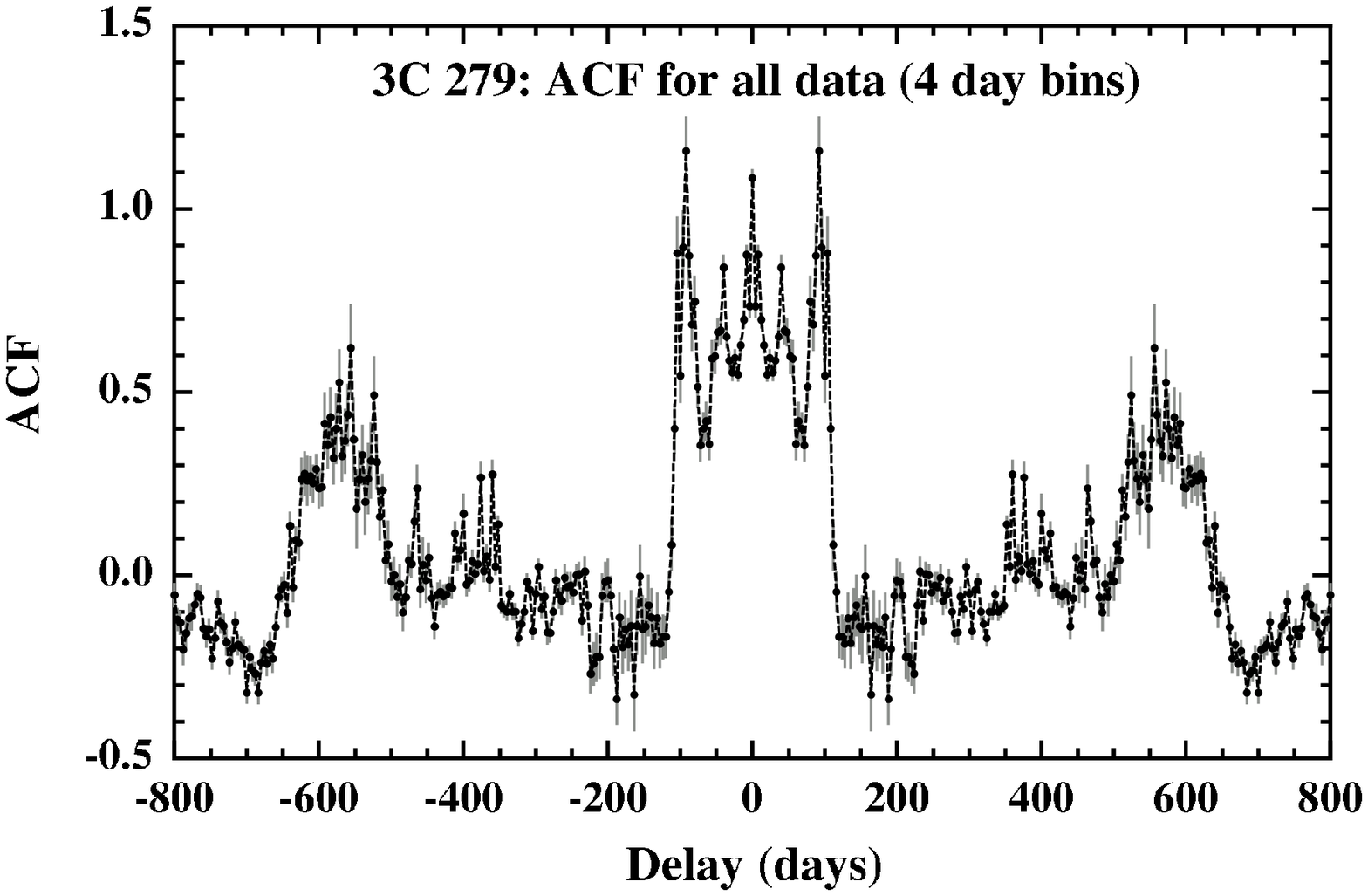}
\caption{The autocorrelation function for the entire 14-year dataset of 3C 279 in 4 day bins. The width of the peak at a time delay of zero days is eight days and there are peaks at 550, 360, and 100 days.}
\end{figure}

%figure13
\begin{figure}
\epsscale{1.00}
\plotone{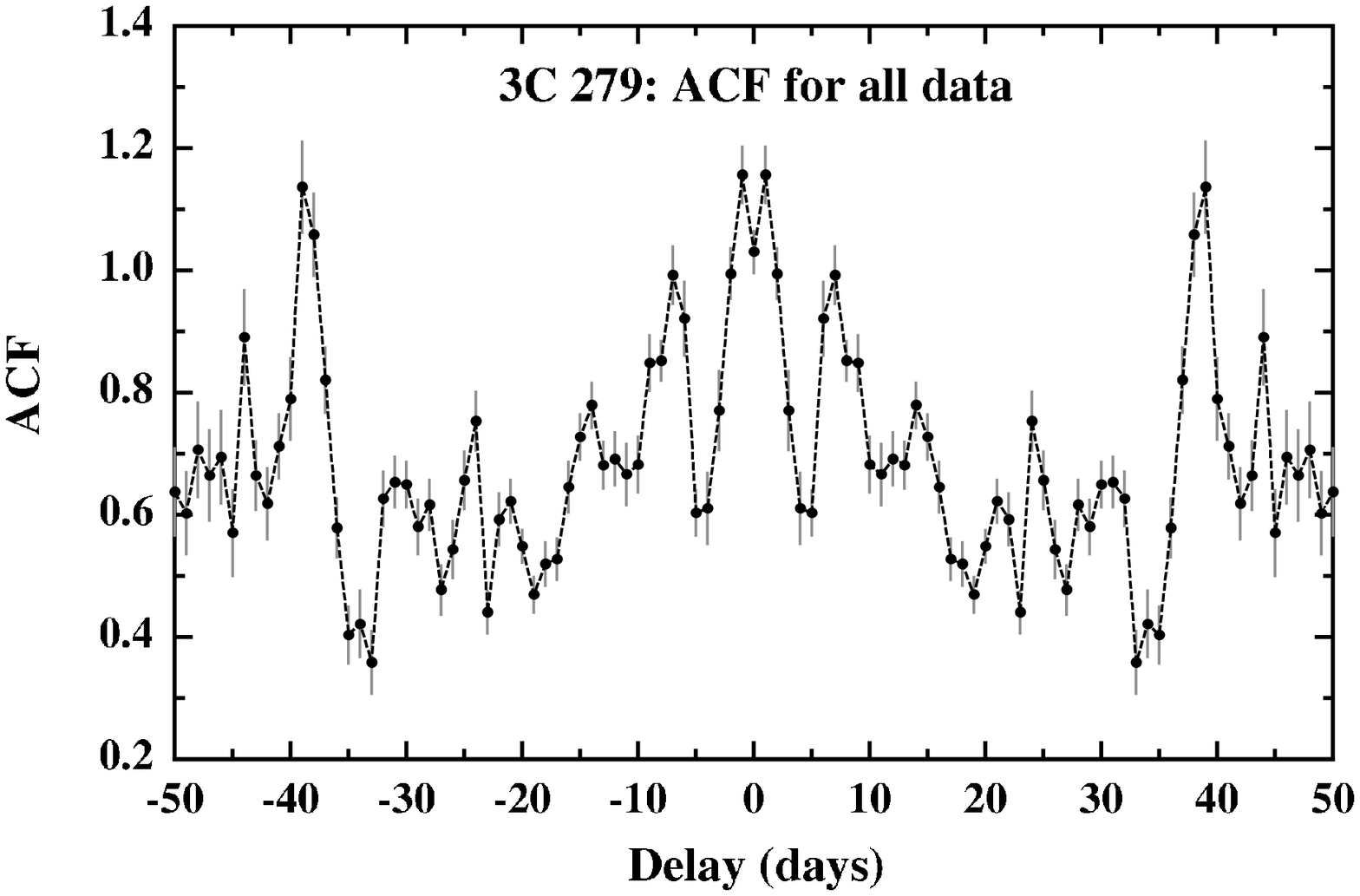}
\caption{The autocorrelation function for the entire 14-year dataset of 3C 279 in 1 day bins. It has higher resolution than Fig 12 and shows peaks at 40 and 7 days.}
\end{figure}

%figure14
\begin{figure}
\epsscale{1.00}
\plotone{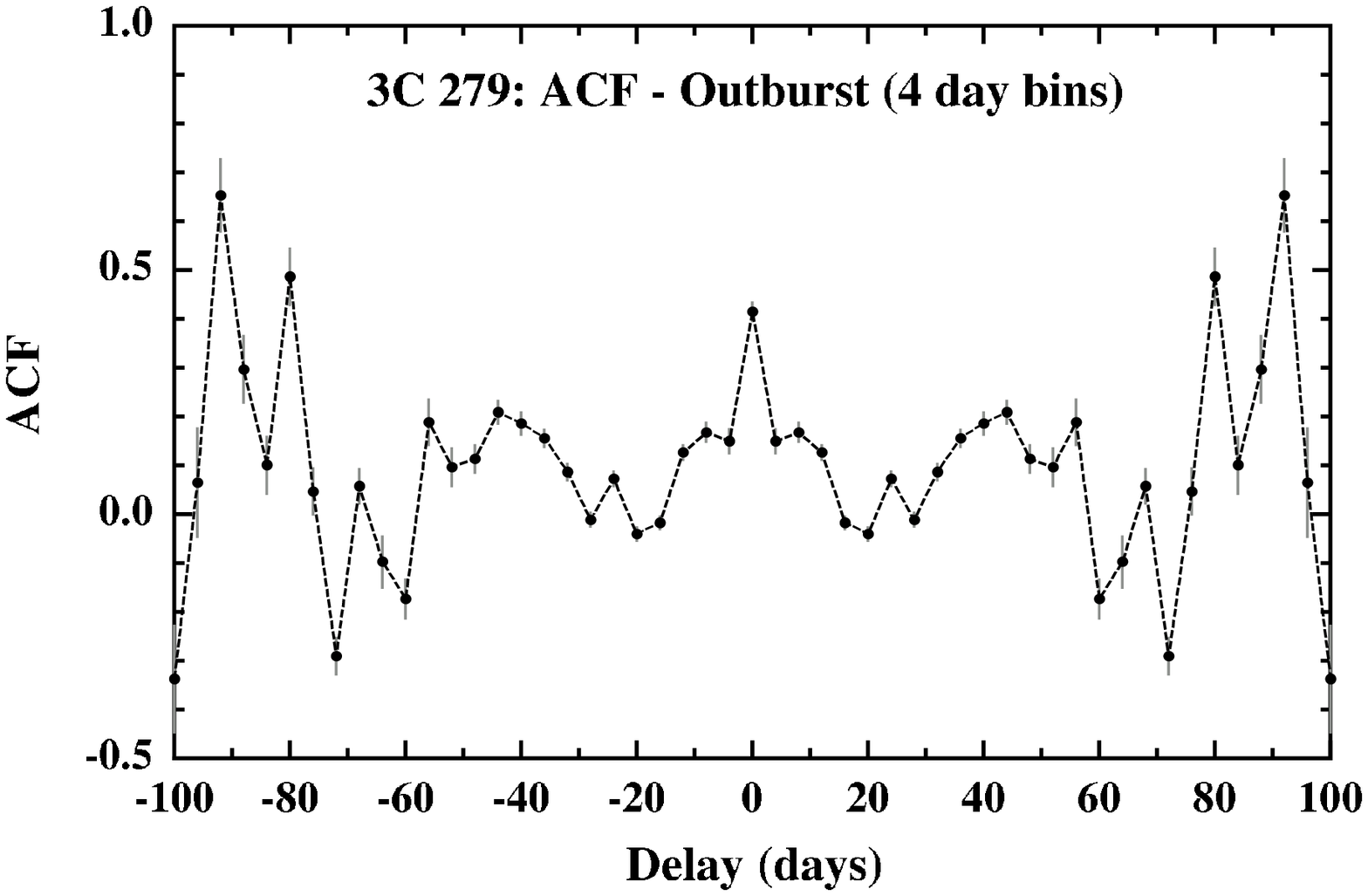}
\caption{The autocorrelation function for the 2001-2002 outburst of 3C 279 in 4 day bins. It has a width of eight days and shows a peak at 45 days.}
\end{figure}

%figure15
\begin{figure}
\epsscale{1.00}
\plotone{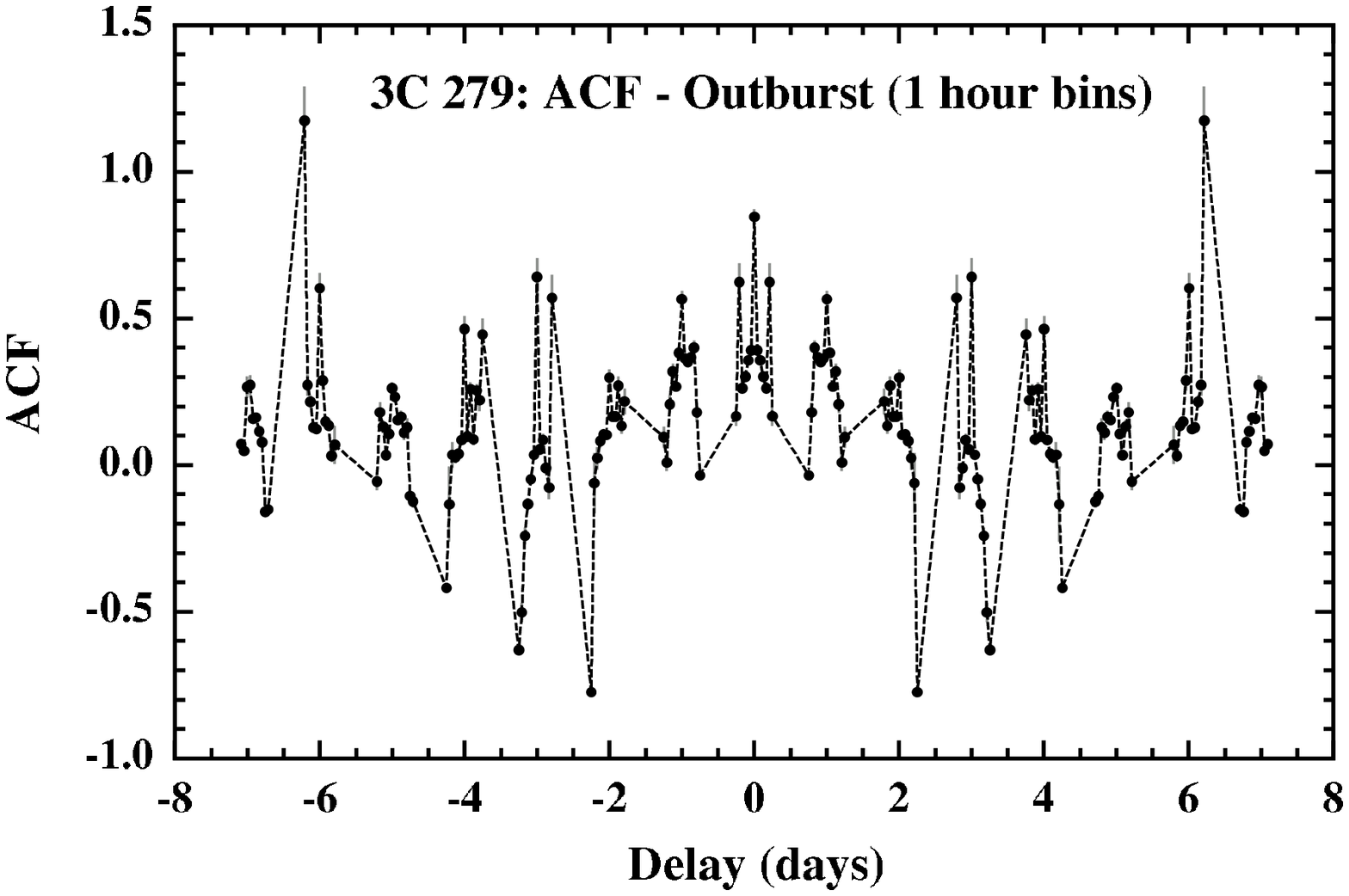}
\caption{The autocorrelation function for the 2001-2002 outburst of 3C 279 in 1 hour bins. It has a width of eight hours and shows a peak at 5 hours as well as pairs of peaks at one-day increments.}
\end{figure}

\clearpage
%figure16
\begin{figure}
\epsscale{1.00}
\plotone{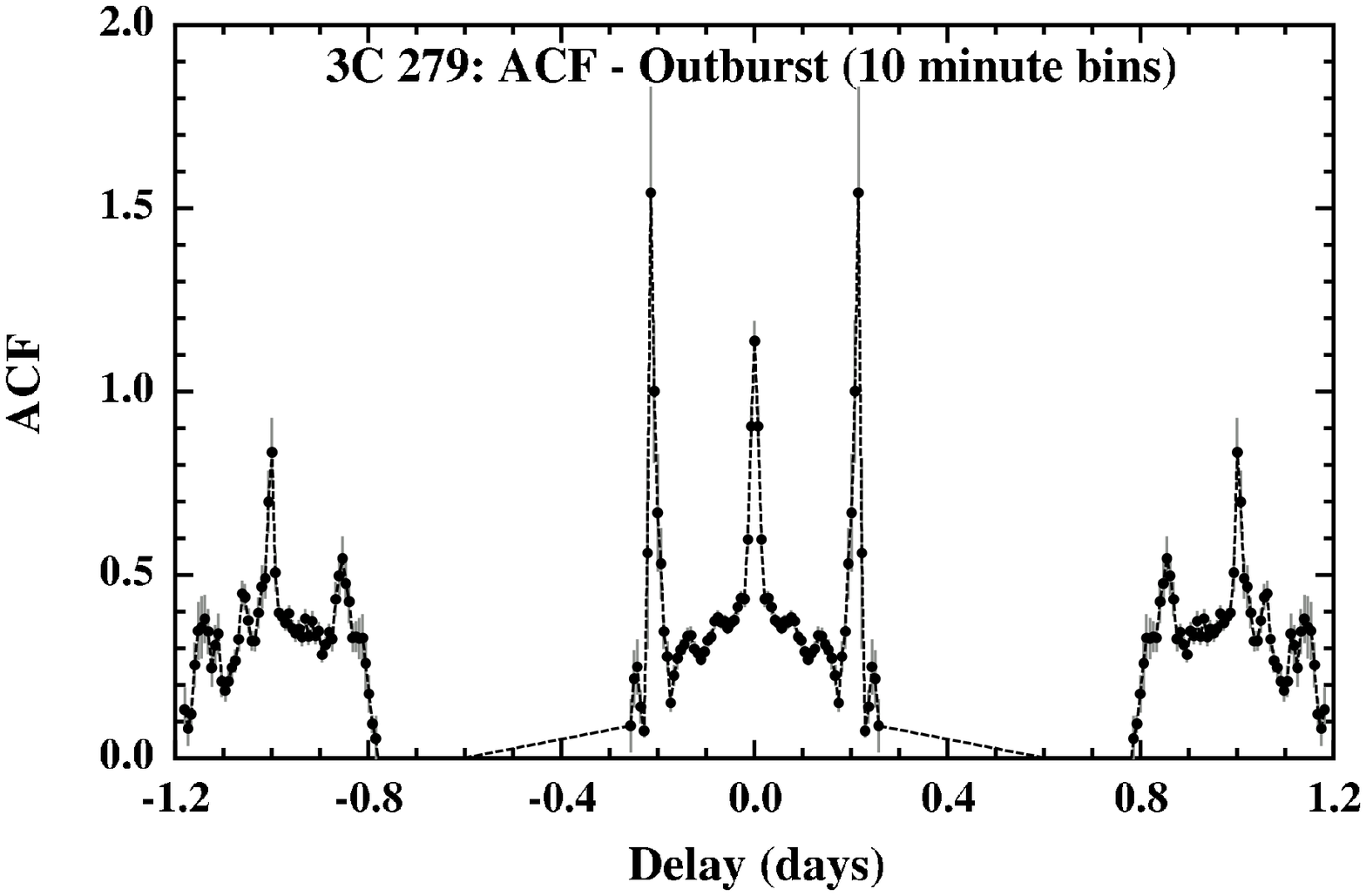}
\caption{The autocorrelation function for the 2001-2002 outburst of 3C 279 in ten minute bins. The width of the function is one hour and it has peaks at 1 day, 20 hours and 5.2 hours.}
\end{figure}

%figure17
\begin{figure}
\epsscale{1.00}
\plotone{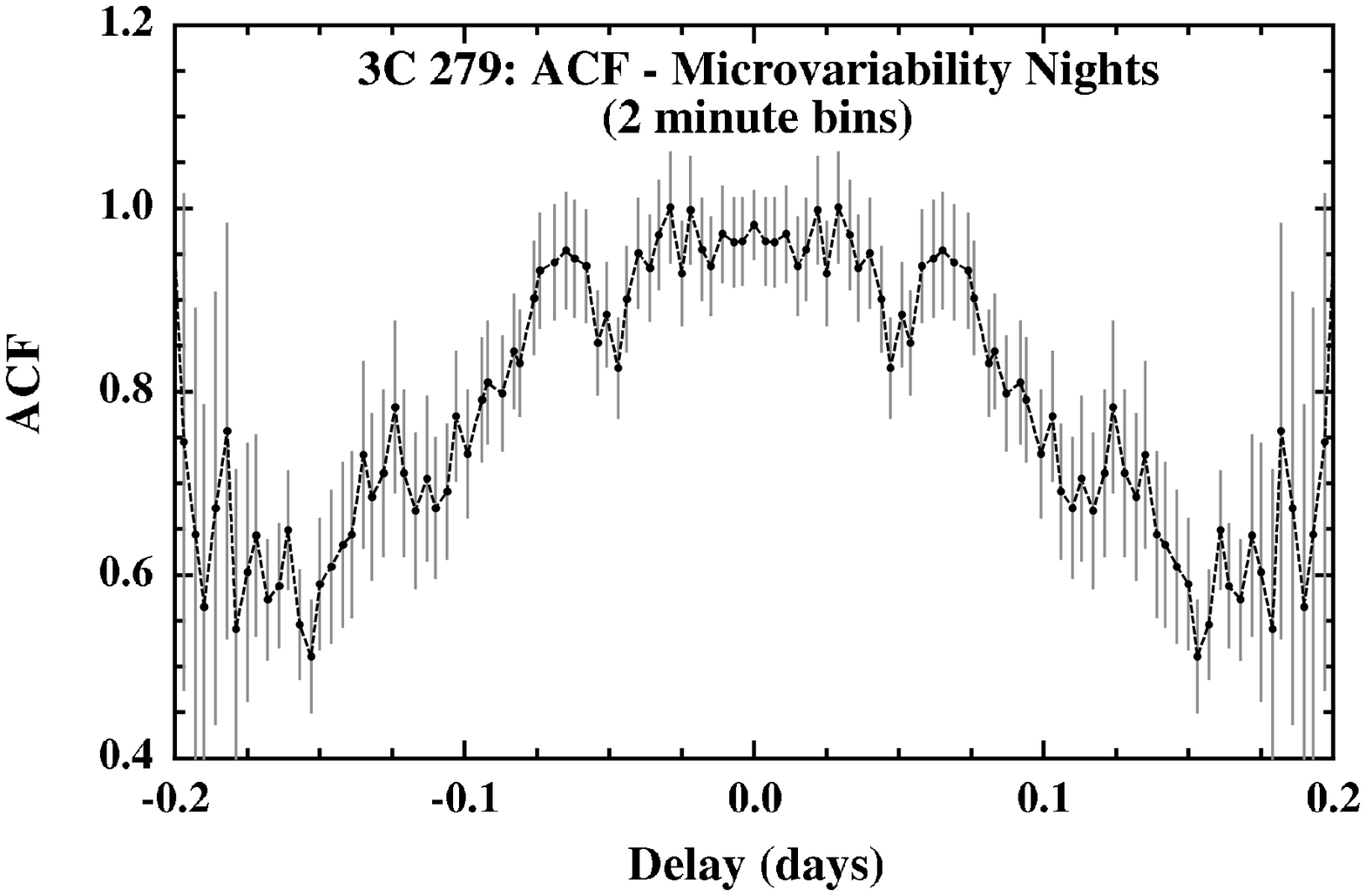}
\caption{The autocorrelation function for the 2001-2002 outburst of 3C 279 in two minute bins. The width is 2.3 hours and there are peaks at 1.5, 3, 5 and 6 hours.}
\end{figure}

%figure18
\begin{figure}
\epsscale{1.00}
\plotone{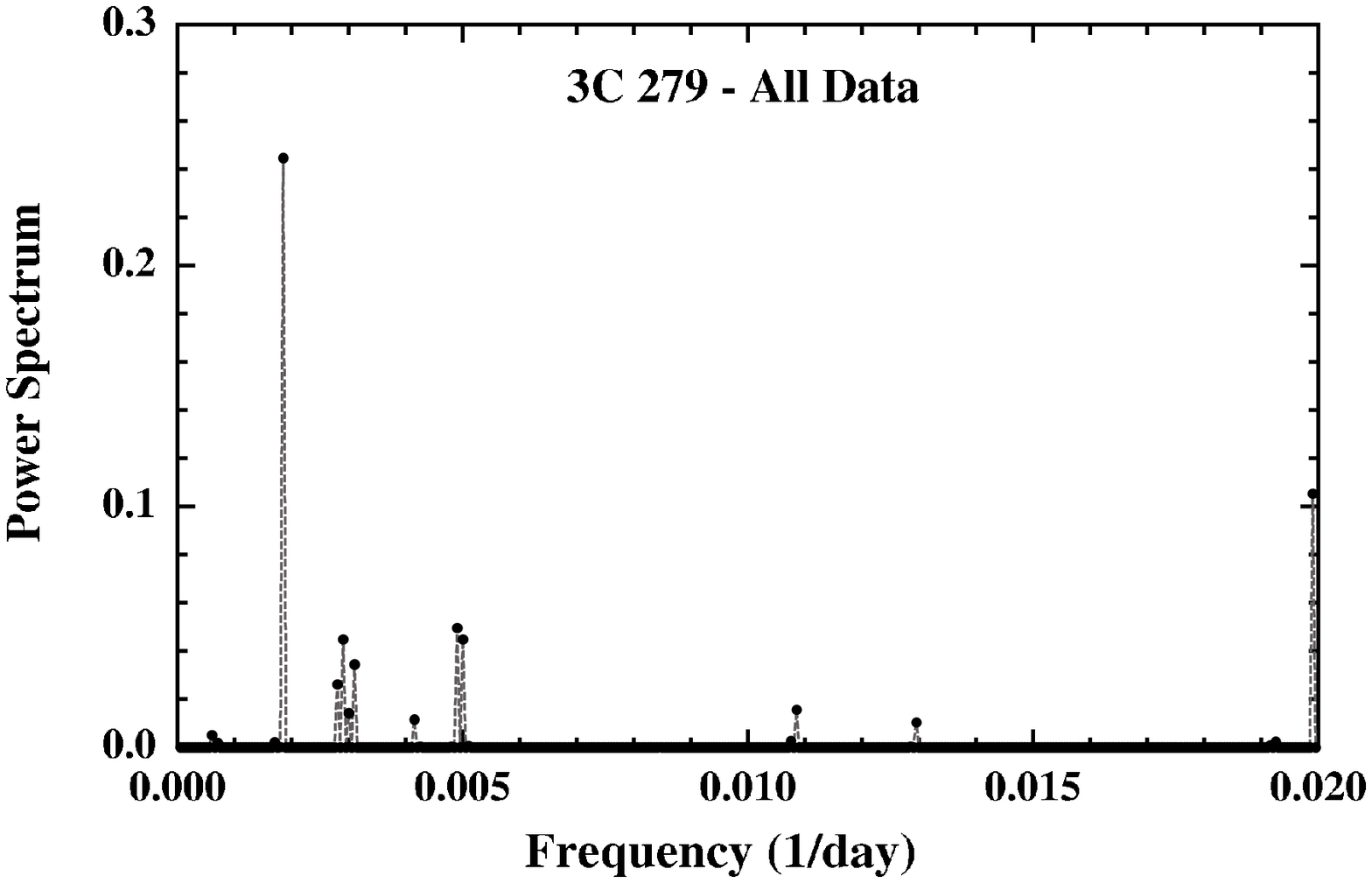}
\caption{A power spectrum for the entire fourteen-year dataset of 3C 279. The clumps of peaks shown are at 540, 345, 200, 100, and 50 days.}
\end{figure}

%figure19
\begin{figure}
\epsscale{1.00}
\plotone{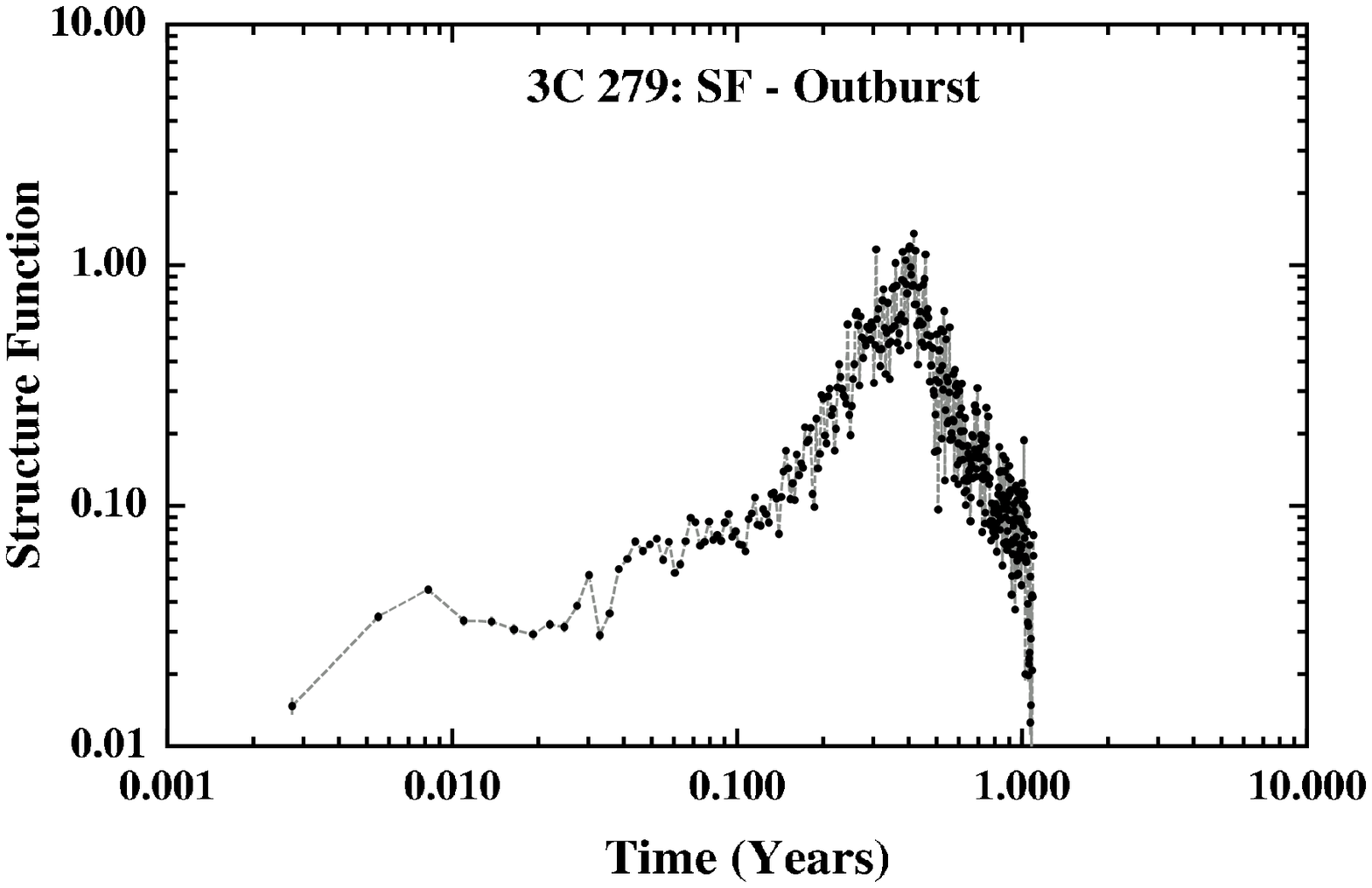}
\caption{The structure function for the 2001-2002 Outburst of 3C 279. The turnover is at 0.4 years and there are possible breaks in the SF at $\sim 46$ and 14 days.}
\end{figure}

%figure20
\begin{figure}
\epsscale{1.00}
\plotone{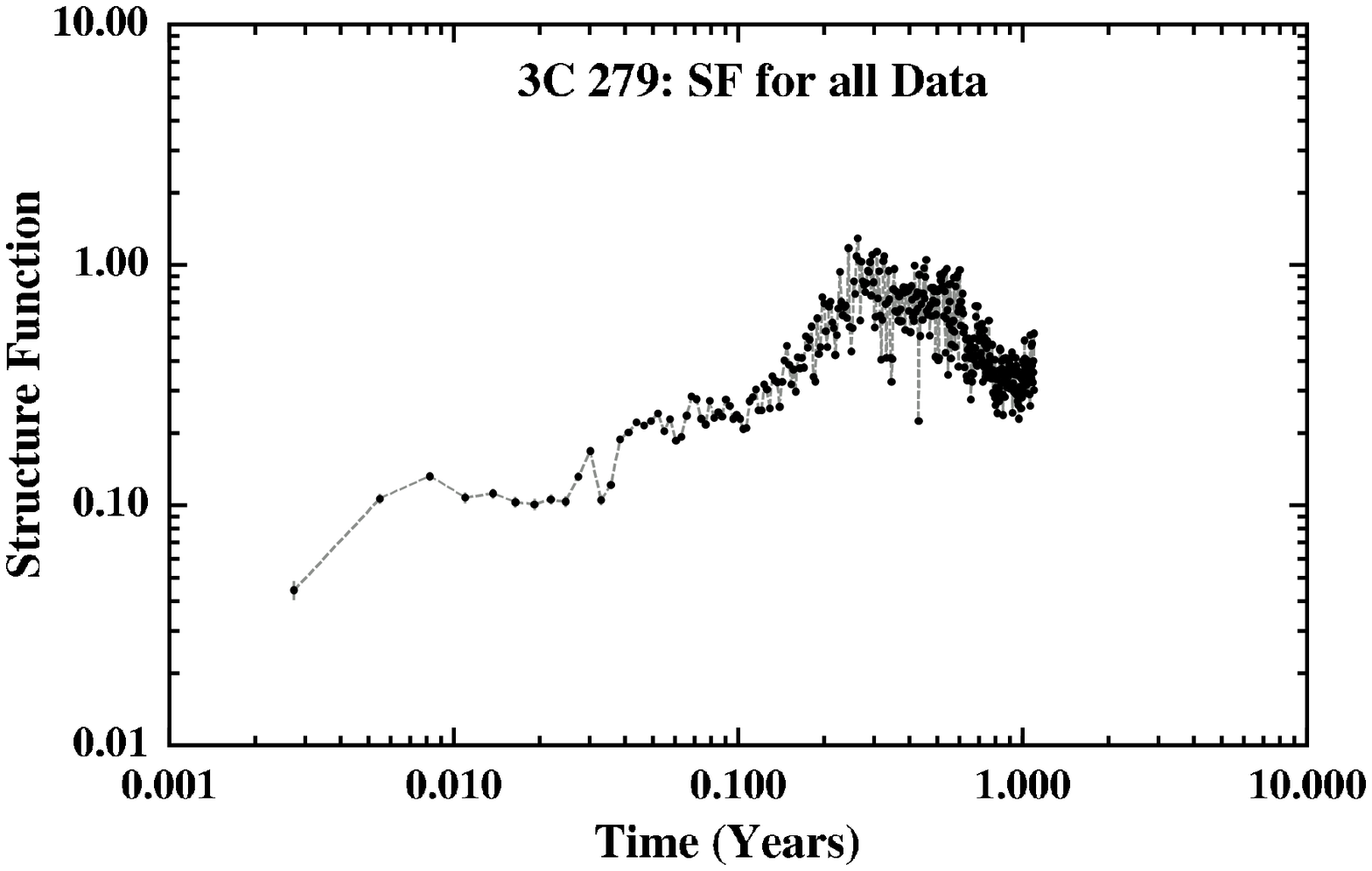}
\caption{The structure function for the entire fourteen-year dataset for 3C 279. The turnover of the SF is at 0.3 years and there are possible breaks at $\sim40$ and 14 days.}
\end{figure}

%figure21
\begin{figure}
\epsscale{1.00}
\plotone{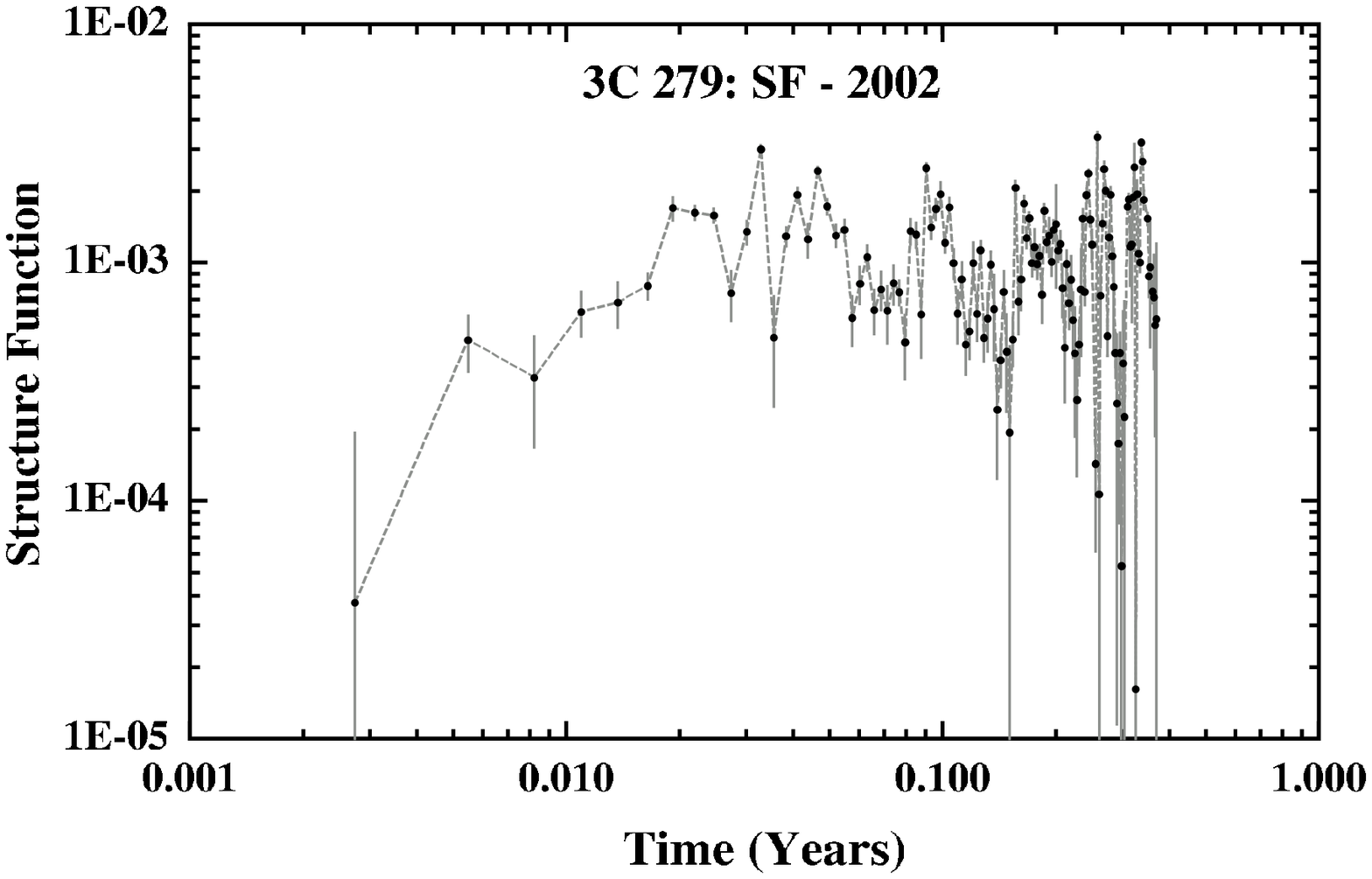}
\caption{The structure function for the 2002 data for 3C 279. There is a break in the SF at 7 days and possibly one at 41 days as well.}
\end{figure}

\clearpage

\begin{deluxetable}{cccc}
  \tablewidth{0pt}
  \tablecaption{Microvariability Coverage Measurements\tablenotemark{a} \label{Microvariability}}
  \tablehead{\colhead{Date (2001)} & \colhead{Time (UT)} & \colhead{Magnitude (R)} &  \colhead{Error}}
    \startdata
    15-Apr & 4.6981 & 14.301 & 0.021 \\
           & 4.7917 & 14.248 & 0.023 \\
           & 4.8378 & 14.266 & 0.020 \\
           & 4.9178 & 14.269 & 0.022 \\
           & 4.9817 & 14.280 & 0.023 \\
    \enddata
    \tablenotetext{a}{The complete version of this table is in the electronic edition of the Journal. The printed edition contains only a sample.}
\end{deluxetable}

\begin{deluxetable}{llcl}
  \tablewidth{0pt}
  \tablecaption{Typical Timescales Found By the Various Numerical Methods \label{Timescales}}
  \tablehead{\colhead{Timescale} & \colhead{Method(s)} & \colhead{CF \tablenotemark{a}} & \colhead{Characteristic of:} }
    \startdata
    1.5 years    & ACF, PS     & 2       & Outburst        \\
    1 year       & ACF, PS     & \nodata & Alias due to sampling interval \\
    100 days     & ACF, PS, SF & 2       & Blaze           \\
    40 - 50 days & ACF, PS, SF & 3       & Flare Ensemble  \\
    7 - 10 days  & ACF, PS, SF & 3       & Flare           \\
    1 day        & ACF, PS     & \nodata & Alias due to sampling interval \\
    8 hours      & ACF         & \nodata & Alias due to data sampling     \\
    5 hours      & ACF         & \nodata & Alias due to data sampling     \\
    1.5 hours    & ACF         & 1       & Microvariability               \\
    2.3 hours    & ACF         & 1       & Microvariability               \\
    \enddata
        \tablenotetext{a}{Confidence Factor representing the reliability of each variability timescale. A CF of 3 means that timescale is the most reliable.}

\end{deluxetable}

\begin{deluxetable}{lcccccc}
  \tablewidth{0pt}
  \setlength{\tabcolsep}{0.05in}
  \tabletypesize{\footnotesize}
  \tablecaption{Summary of the Microvariability Data \label{summary}}
  \tablehead{\colhead{UT Date} & \colhead{Hours} & \colhead{Average} & \colhead{Weather} & \colhead{Overall Slope} & \colhead{$1^{st}$ Slope} & \colhead{$2^{nd}$ Slope} \cr  
  \colhead{(2001)} & \colhead{Observed} & \colhead{Error Bar} & \colhead{Conditions} & \colhead{(mag/hr)} & \colhead{(early)} & \colhead{(later)}}
    \startdata
		15-Apr & 3.4 & 0.029 & Clear        &  0.001 & \nodata & \nodata \\
		19-Apr & 6.0 & 0.017 & Clear        &  0.019 & -0.016  &  0.032  \\
		20-Apr & 6.3 & 0.013 & Clear        &  0.005 &  0.017  & -0.007  \\
		27-Apr & 6.2 & 0.019 & Clear        &  0.011 & \nodata & \nodata \\
		28-Apr & 5.0 & 0.016 & Mostly Clear &  0.000 & \nodata & \nodata \\
		29-Apr & 3.4 & 0.010 & Clear        & -0.003 & \nodata & -0.010  \\
		30-Apr & 4.0 & 0.010 & Hazy         &  0.005 & 0.020   &  0.005  \\
		1-May  & 4.7 & 0.012 & Mostly Clear & -0.005 & \nodata & \nodata \\
		2-May  & 2.3 & 0.020 & Variable     &  0.004 & \nodata & \nodata \\
		8-May  & 4.7 & 0.017 & Clear        & -0.016 & \nodata & \nodata \\
		10-May & 2.3 & 0.020 & Clear        &  0.014 & \nodata & \nodata \\
		11-May & 2.0 & 0.016 & Clear        & -0.009 & \nodata & \nodata \\
		14-May & 3.7 & 0.010 & Clear        & -0.001 & \nodata & \nodata \\
		15-May & 4.0 & 0.010 & Clear        & -0.032 & \nodata & \nodata \\
		16-May & 1.3 & 0.011 & Clear        &  0.038 & \nodata & 0.048   \\
		17-May & 2.6 & 0.023 & Variable     & -0.031 & \nodata & \nodata \\
		20-May & 3.7 & 0.016 & Clear        & -0.025 & \nodata & \nodata \\
		21-May & 2.6 & 0.032 & Variable     & -0.008 & \nodata & \nodata \\
		25-May & 1.9 & 0.015 & Variable     & -0.001 & \nodata & \nodata \\
		28-May & 1.2 & 0.013 & Clear        &  0.006 & \nodata & \nodata \\
		30-May & 2.6 & 0.016 & Clear        & -0.009 & \nodata & \nodata \\
		31-May & 3.0 & 0.020 & Clear        &  0.000 & \nodata & \nodata \\
		7-Jun  & 2.3 & 0.011 & Clear        &  0.012 & \nodata & \nodata \\
		8-Jun  & 1.5 & 0.007 & Mostly Clear & -0.030 & 0.017   & \nodata \\
		9-Jun  & 2.1 & 0.009 & Clear        & -0.005 & \nodata & \nodata \\
		13-Jun & 1.1 & 0.016 & Variable     &  0.004 & \nodata & \nodata \\
		14-Jun & 1.3 & 0.015 & Clear        & -0.004 & -0.048  & 0.056   \\
		18-Jun & 1.5 & 0.014 & Clear        & -0.009 & \nodata & \nodata \\
		20-Jun & 1.3 & 0.016 & Clear        &  0.040 & \nodata & \nodata \\
    \enddata
\end{deluxetable}

 \end{document}